\documentclass[12pt]{article}
\usepackage{amssymb,amsmath,epsfig}
\allowdisplaybreaks
\begin{document}
\title{\bf Static Wormhole Solutions and Noether Symmetry in Modified Gauss-Bonnet Gravity}
\author{M. Sharif \thanks{msharif.math@pu.edu.pk}~, Iqra Nawazish
\thanks{iqranawazish07@gmail.com} and Shahid Hussain
\thanks{shahidhussaine56@gmail.com}\\
Department of Mathematics, University of the Punjab,\\
Quaid-e-Azam Campus, Lahore-54590, Pakistan.}
\date{}

\maketitle

\begin{abstract}
In this paper, we analyze static traversable wormholes via Noether
symmetry technique in modified Gauss-Bonnet $f(\mathcal{G})$ theory
of gravity (where $\mathcal{G}$ represents Gauss-Bonnet term). We
assume isotropic matter configuration and spherically symmetric
metric. We construct three $f(\mathcal{G})$ models, i.e, linear,
quadratic and exponential forms and examine the consistency of these
models. The traversable nature of wormhole solutions is discussed
via null energy bound of the effective stress-energy tensor while
physical behavior is studied through standard energy bounds of
isotropic fluid. We also discuss the stability of these wormholes
inside the wormhole throat and conclude the presence of traversable
and physically stable wormholes for quadratic as well as exponential
$f(\mathcal{G})$ models.
\end{abstract}

\section{Introduction}

The general theory of relativity (GR) not only incorporates
information about gravity and matter but also provides foundation
for the understanding of black holes and standard big-bang model of
cosmology. GR is the simplest relativistic
theory of gravity that is consistent with the experimental data but
still suffers from some unresolved issues like earlier and current
cosmic expansions. The favorable and optimistic approach to unveil
the salient features of these dark aspects is to modify the gravity
by introducing some extra degrees of freedom in the Einstein-Hilbert
action. These modifications are formulated by replacing or adding
curvature invariants as well as their corresponding generic
functions in Einstein-Hilbert action referred as modified gravitational theories
\cite{1a}.

Recent observational facts of modern cosmology indicate the current
accelerated expansion of the universe. This expansion occurs due to
the strange force with fascinating anti-gravitational impacts, named
as dark energy (DE). One of the approaches to study the nature of DE
is the modified theories of gravity. Nojiri and Odintsov
\cite{1} proposed $f(\mathcal{G})$ ($\mathcal{G}$ represents
Gauss-Bonnet (GB) invariant) gravity by including higher-order
correction terms. The inspiration of this theory arises from the
string theory at low energy scale which efficiently helps to examine
the late-time evolution of the cosmos. The GB invariant is quadratic
in nature and is free from spin-2 ghost instabilities acts as a
four-dimensional topological term which is the composition of the
scalar curvature ($R$), Ricci ($R_{\alpha\beta}$) and Riemann
tensors ($R_{\alpha\beta\mu\nu}$) defined as
$\mathcal{G}=R^{2}-4R_{\alpha\beta}R^{\alpha\beta}
+R_{\alpha\beta\mu\nu}R^{\alpha\beta\mu\nu}$.

This theory has a quite rich cosmological structure which describes
fascinating characteristics of early as well as late-time
cosmological evolution and is consistent with solar system
constraints. The GB invariant gives fascinating results when either
comprised of a scalar field or a general function $f(\mathcal{G})$
is included in the Einstein-Hilbert action \cite{2}-\cite{4}. This
theory provides a possibility to study the transformation from
non-phantom to phantom phase and from decelerated to an accelerated
region. It is observed that $f(\mathcal{G})$ gravity well describes
the laws of thermodynamics and many other cosmological issues
\cite{5}-\cite{8}. Sharif and Fatima \cite{9} investigated the
spherical interior solutions of this gravity by applying conformal
Killing vectors corresponding to isotropic as well as anisotropic
fluid configurations and checked the physical consistency via energy
conditions.

Noether symmetry is recognized as the most efficient method to
investigate the analytic solutions that help to find the conserved
parameters of the field equations corresponding to symmetry
generators. Capozziello et al. \cite{10} examined the analytic
solutions of static spherically symmetric spacetime for the
power-law functional form of $f(R)$ theory. The same authors
\cite{11} extended this work for the non-static case and obtained
exact solutions for constant as well as variable curvature scalar.
Vakili \cite{12} used this approach for flat FRW model to discuss
the current cosmic expansion through an effective equation of state (EoS)
parameter in $f(R)$ gravity. Many researchers \cite{13} investigated
the current accelerated cosmic expansion through this approach in
different modified theories.

A wormhole (WH) is a hypothetical bridge or tunnel that allows a
smooth passing through different regions of spacetime. If
hypothetical tunnel connects two regions of the same spacetime then
intra-universe WH is established whereas inter-universe WH appears
for two distinct spacetimes. The existence of exotic matter (matter
with negative energy density) encourages observer to move smoothly
through tunnel but its sufficient amount leads to controversial
existence of a realistic WH. Consequently, the only way to have a
physically viable WH model is to minimize the usage of exotic matter
in the tunnel. For any static configuration, the most crucial
problem is stability analysis which defines their behavior against
perturbations as well as enhances physical characterization. A
singularity free configuration identifies a stable state which
successfully prevents the WH to collapse while a WH can also exist
for quite a long time even if it is unstable due to very slow decay.

The study of WH geometries has gained much attention in modified
theories of gravity. In $f(R)$ scenario, Lobo and Oliveira \cite{14}
assumed distinct fluid distributions with constant shape function to
investigate the WH geometry. Jamil et al. \cite{15} examined
feasible WH solutions with non-commutative geometry by considering a
specific shape function corresponding to power-law $f(R)$ model.
Bahamonde et al. \cite{16} used the same gravity for FRW universe
model to analyze the cosmological WH solutions with isotropic fluid.
Mazharimousavi and Halilsoy \cite{17} discussed the conditions of WH
for vacuum/non-vacuum cases and obtained the stable WH geometry for
$f(R)$ model along with polynomial evolution. Sharif and Fatima
\cite{18} explored the non-static solutions of WH as well as static
spherically symmetric WH in galactic halo region in $f(\mathcal{G})$
gravity. Bahamonde et al. \cite{19} found definite solutions of
shape function and red-shift parameter via Noether symmetry and
examined the graphical behavior in the background of non-minimal
coupling with torsion scalar in scalar-tensor theory.

Recently, Sharif and Nawazish investigated the static WH solutions
using Noether symmetry technique in both $f(R)$ \cite{20} as well as
$f(R,T)$ gravity \cite{21} and found stable structure for two
different values of red-shift function. In this paper, we study the
physical presence of WH via Noether symmetry technique in
$f(\mathcal{G})$ theory and explore WH properties associated with
perfect fluid. The paper is arranged in the following pattern.
Section \textbf{2} represents the basic formalism of this gravity.
We obtain point-like Lagrangian in section \textbf{3} which is used
in section \textbf{4} to estimate WH solutions for variable
red-shift function. Section \textbf{5} explores the stable structure
of developed WH geometries and summary of our results is given in
the last section.

\section{Basic Formalism of $f(\mathcal{G})$ Gravity}

The action of $f(\mathcal{G})$ gravity in 4-dimensions with matter
Lagrangian is presented by
\begin{equation}\label{1}
S=\frac{1}{2k^2}\int
\left[R+f\left(\mathcal{G}\right)\right]\sqrt{-g}d^4x+\int
\sqrt{-g}\mathcal{L}_m d^4x,
\end{equation}
where $k$ is the coupling constant and $\mathcal L_m$ defines matter
Lagrangian. Varying this action with respect to metric tensor, the
corresponding field equations are
\begin{eqnarray}\nonumber
G_{\alpha\beta}&=&\frac{1}{2}g_{\alpha\beta}
f(\mathcal{G})-(2RR_{\alpha\beta}-4R^{\mu}_{\alpha}R_{\mu\beta}-
4R_{\alpha\mu\beta\nu}R^{\mu\nu}+2R^{\mu\nu\gamma}_{\alpha}
R_{\beta\mu\nu\gamma})f_{\mathcal{G}}\\\nonumber&-&(2R\nabla^2
g_{\alpha\beta}-2\nabla_{\alpha}\nabla_{\beta}R
-4R^{\mu\nu}g_{\alpha\beta}\nabla_\mu\nabla_\nu-4\nabla^2R_{\alpha\beta}
+4\nabla_{\beta}\nabla_{\mu}R^{\mu}_{\alpha}\\\label{2}&+&4
\nabla_{\alpha}\nabla_{\mu}R^{\mu}_{\beta}
+4\nabla^{\mu}\nabla^{\nu}R_{\alpha\mu\beta\nu})
f_\mathcal{G}+k^2T_{\alpha\beta},
\end{eqnarray}
where $\nabla^{2}=\nabla_{\alpha}\nabla^{\alpha}$ is d'Alembert
operator, $\nabla_{\alpha}$ indicates the covariant derivative and
$f_{\mathcal{G}}$ denotes differentiation of generic function with
respect to $\mathcal{G}$. The stress-energy tensor is determined by
the following form
\begin{equation}\label{3}
T_{\alpha\beta}=\frac{-2}{\sqrt{-g}}\frac{\delta\left(\sqrt{-g}
\mathcal L_m\right)} {\delta\left(g^{\alpha\beta}\right)}.
\end{equation}
Here the metric tensor depends only upon the distribution of matter
yielding
\begin{equation}\label{4}
T_{\alpha\beta}=g_{\alpha\beta}\mathcal L_m-2\frac{\delta \mathcal
L_m}{\delta g^{\alpha\beta}}.
\end{equation}
The energy-momentum tensor for perfect fluid configuration is
\begin{equation}\label{2a}
T_{\alpha\beta}^{(m)}=(\rho_m+p_m)u_{\alpha}u_{\beta}
+p_mg_{\alpha\beta},
\end{equation}
where $p_m$ and $\rho_m$ characterize pressure and energy density,
respectively and $u_{\alpha}$ represents the four velocity of the
fluid.

The static spherically symmetric line element \cite{22} is given by
\begin{equation}\label{5}
ds^{2}=-e^{a(r)}dt^{2}+e^{b(r)}dr^{2}+M(r)(d\theta^{2}+\sin^{2}\theta
d\phi^{2}),
\end{equation}
where the triplet $(M,a,b)$ indicates generic radial functions. For
$M(r)=r^2$ \cite{23}, the spherical symmetry (\ref{5}) characterizes
Morris-Thorne WH in which $a(r)$ is identified as the red-shift
function as it determines gravitational red-shift of WH whereas
$e^{b(r)}=(1-\frac{h(r)}{r})^{-1}$ with $h(r)$ being the shape
function as it specifies spacial shape of WH. The radial coordinate
$r$ possesses non-monotonic behavior as it decreases from infinity
to minimum radius when a traveler moves from one part of WH. The
space occupying minimum radius is known as throat of WH. The radial
coordinate starts increasing from minimum radius to infinity as
traveler comes out of the throat and entered into another region of
WH. The basic property of WH is the flaring-out condition for which
$\frac{h(r)-h(r)'r}{h(r)^2}>0$. At the throat or near the throat,
the traversable WH demands $0\leq h'(r)<1$, where prime represents
differentiation with respect to $r$. The sufficient condition of
traversable WH is the finite red-shift function throughout the whole
space of WH. This condition ensures the absence of horizons and
consequently, allows a traveler to move into a WH as well as
appreciates a smooth exit.

For spherically symmetric spacetime (5) and perfect fluid (6), we
formulate the field equations corresponding to Eqs.(\ref{1}) and
(\ref{2}) as follows
\begin{eqnarray}\nonumber
&&\frac{e^a\left(-4M''M+2b'M'M+M'^2+4Me^b\right)}{4e^bM^2}=
\rho_me^ak^2-\frac{1}{2}e^af\left(\mathcal{G}\right)
\\\nonumber&+&\left.e^{a-2b}\left[a'^4-\frac{M'a'b'}{2M^2}
+\frac{M'a'b'^2}{M}-\frac{3M'^2a'b'}{4M^2}-2\frac{M'a''b'}{M}
+4\frac{M''a''}{M}-\frac{e^ba'^2}{M}\right.\right.\\\nonumber&
+&\left.\left.\frac{e^ba'b'}{M}-3\frac{e^ba'M'}{M^2}
-2\frac{e^ba''}{M}+2\frac{M''a'^2}{M}-2\frac{M''a'b'}{M}
+3\frac{M''a'M'}{M^2}-\frac{3b'a'^3}{4}\right.\right.\\
\nonumber&-&\left.\left.\frac{a'M'^3}{2M^3}
-\frac{3a''M'^2}{2M^2}+2a''^2-\frac{a'^2b'M'}{2M}
+2a''a'^2-\frac{3a''a'b'}{2}+\frac{a'^2b'^2}{8}\right]
f_\mathcal{G}\right.\\\nonumber&-&\left.e^{a-2b}
\left(\frac{4a''M'}{M}-\frac{b'M'^2}{M^2}+4\frac{M'M''}
{M^2}-\frac{M'^3}{M^3}-\frac{a'^3}{2}+\frac{3a'^2b'}{2}
-a'a''+2a''b'\right.\right.\\\nonumber&+&\left.\left.
\frac{a'b'M'}{M}-\frac{a'^2M'}{M}+2\frac{a'M'^2}{M^2}
-a'b'\right)-e^{a-b}\left(a'^3-b'a'^2+2\frac{M'a'^2}{M}
+2a''a'\right.\right.\\\nonumber&-&\left.\left.e^{2
a-2b}a'b'^2-2\frac{M'^2}{M^2}\right)f_\mathcal{G}'
+\left[\frac{4}{M}e^{a-b}+e^{a-2b}\left(\frac{M'^2}{M^2}
+4\frac{M''}{M}+3a'^2-4a'b'\right.\right.\right.\\\label{6}&+&
\left.\left.5a''\right)\right]f_\mathcal{G}'', \\\nonumber
&&-\frac{\left(M'^2+2a'M'M+M'^2-4Me^b\right)}{4M^2}=k^2p_me^b
+2p_me^b+\frac{1}{2}e^bf\left(\mathcal{G}\right)\\\nonumber&-&
\left.e^{-b}\left[\left(\frac{a'b'^2M'}{2M}+\frac{a'^2M'^2}{4M^2}
-7\frac{a'b'M'^2}{M^2}+\frac{a'^3M'}{M}+2\frac{a'M'a''}{M}
+2\frac{a'M'M''}{M}\right.\right.\right.\\\nonumber&-&\left.
\left.\left.\frac{3a'M'^3}{2M^3}+\frac{11b'M'^3}{4M^3}
+\frac{a''M'^2}{2M^2}-\frac{a'^2e^b}{M}+\frac{a'b'e^b}{M^2}
+\frac{4b'M'e^b}{M^2}-\frac{2a''e^b}{M}\right.\right.\right.
\\\nonumber&-&\left.\left.\left.\frac{4M''e^b}{M^2}
+\frac{4M'^2e^b}{M^3}+\frac{4M''^2}{M^2}-\frac{4M''M'^2}{M^3}
-\frac{M'^4}{2M^4}-\frac{a'b'^2M'}{2M}-\frac{2b'M'^3}{M^2}
\right.\right.\right.\\\nonumber&+&\left.\left.\left.\frac{4M'^3M''}
{M^3}+\frac{a'^4}{4}-\frac{a'^3b'}{2}-a'b'a''+a''^2+\frac{a'^2b'^2}
{4}+\frac{b'^2M'^2}{M^2}-\frac{4b'M'M''}{M^2}\right.\right.\right.
\\\nonumber&+&\left.\left.\left.\frac{a'M'M''}{M^2}-\frac{4M''e^b}{M^2}
-2M''M'^2\right)f_\mathcal{G}+\frac{a'^3}{2}+a'^2M+a''a'M
+\left(\frac{2M'^3}{M^3}\right.\right.\right.\\\nonumber&-&\left.
\left.\left.\frac{a'^2M'}{M}-\frac{5a'M'^2}{2M^2}+\frac{3b'M'^2}{2}
+\frac{2a'e^b}{M}+\frac{4M'e^b}{M^2}-\frac{2M'M''}{M^2}-\frac{2b'M''}
{M}\right.\right.\right.\\\label{7}&+&\left.\left.\frac{4MM''}{M^2}
-\frac{a'^3}{2}-a'a''\right)f_\mathcal{G}'\right], \\\nonumber
&&\frac{M'M\left(a'-b'\right)+2M''M+M^2a'^2-M^2a'b'-M'^2+2M^2a''}{4
Me^b}=k^2p_mM\\\nonumber&+&\left.\frac{1}{2}f\left(\mathcal{G}\right)
-e^{-2b}\left(\frac{a'^3M'}{2}-\frac{3a'^2b'M'}{4}-a'^2e^b+\frac{a'^2M''}
{2}+\frac{a'b'^2M'}{4}+a'b'e^b\right.\right.\\\nonumber&-&\left.
\left.\frac{a'b'M''}{2}+\frac{a'^2M'^2}{M}-\frac{3a'b'M'^2}{4M}
+\frac{a'M'a''}{M}+\frac{b'^2M'^2}{2M}+\frac{2b'M'e^b}{M}
+a''M''\right.\right.\\\nonumber&+&\left.\left.a'M'a''
-\frac{a''b'M'}{2}-2a''e^b-\frac{2b'M'M''}{M}-\frac{2a'M'e^b}{M}
+\frac{8e^{2b}}{M}-\frac{4M''e^b}{M}\right.\right.\\\nonumber&+&\left.
\left.\frac{2M''^2}{M}+\frac{b'M'^3}{2M^2}-\frac{M''M'^2}{M^2}
+\frac{M'^4}{2M^3}-\frac{2M'^2e^b}{M^2}\right)f_\mathcal{G}
-e^{-2b}\left[\frac{a'^3M^2}{2}\right.\right.\\\nonumber&+&\left.a'^2M'M
-\frac{a'^2b'M^2}{2}+a''a'M^2+\left(-\frac{a'^3M}{2}+\frac{3a'b'M'}{2}
-\frac{a'M'^2}{2M}+\frac{13b'M'^2}{8M}\right.\right.\\\nonumber&-&\left.
\left.\left.a''a'M-a''M'+\frac{11M'e^b}{2M}-a'M''+\frac{9a'^2M'}{8}
-\frac{a'^2b'M}{2}+\frac{5b'^2M'}{8}-\frac{4M'}{M}\right.\right.
\right.\\\label{8}&-&\left.\left.\frac{5b'M''}{4}+\frac{M'^3}{8M^2}\right)
f_\mathcal{G}'+\left(a'M'+\frac{5M'^2}{4M}+\frac{5b'M'}{4}
-\frac{5M''}{2}\right)f_\mathcal{G}''\right].
\end{eqnarray}

The energy bounds indicate the nature of matter incorporated by
astrophysical configurations. If the well-defined bounds are
preserved then the configurations are said to be supported by an
ordinary matter. In case of WH geometry, a realistic WH
configuration may exist if these energy bounds violate. In order to
define such energy bounds, Raychaudhari equations are considered to
be the most fundamental ingredients given as
\begin{eqnarray}\label{A}
\frac{d\theta}{d\tau}=-\frac{1}{3}\theta^2-\sigma_{\mu\nu}\sigma^{\mu\nu}
+\Theta_{\mu\nu}\Theta^{\mu\nu}-R_{\mu\nu}l^\mu l^\nu,\\\label{B}
\frac{d\theta}{d\tau}=-\frac{1}{2}\theta^2-\sigma_{\mu\nu}\sigma^{\mu\nu}
+\Theta_{\mu\nu}\Theta^{\mu\nu}-R_{\mu\nu}k^\mu k^\nu,
\end{eqnarray}
where $\theta,~l^\mu,~k^\mu,~\sigma$ and $\Theta$ represent
expansion scalar, timelike vector, null vector, shear and rotation
tensors. These equations are defined for both timelike (first
equation) and null (second equation) congruence. In both equations,
the positivity of last term demands attractive gravity. For the
Einstein-Hilbert action, these energy bounds are split into null
(NEC) ($\rho_{m}+p_{m}\geq0$), weak (WEC)
($\rho_{m}\geq0,~\rho_{m}+p_{m}\geq0$), strong (SEC)
($\rho_{m}+p_{m}\geq0,~\rho_{m}+3p_{m}\geq0$) and dominant (DEC)
($\rho_{m}\geq0,~\rho_{m}\pm p_{m}\geq0$) energy conditions
\cite{I15}. These conditions originate from the Raychaudhari
equations purely on geometric arguments, hence are valid for any
modified theory implying that $T^{(m)}_{\mu\nu}k^\mu k^\nu\geq0$ can
be replaced with $T^{eff}_{\mu\nu}k^\mu k^\nu\geq0$. For detailed
study of energy conditions in modified gravity, see the literature
\cite{15a}. In modified Gauss-Bonnet gravity, these energy
constraints become
\begin{itemize}
\item \textbf{NEC:} \quad$\rho_{eff}+p_{eff}\geq 0$,
\item \textbf{SEC:} \quad$\rho_{eff}+p_{eff}\geq 0,\quad\rho+
3p_{eff}\geq0$, \item \textbf{DEC:} \quad$\rho_{eff}\geq
0,\quad\rho_{eff}\pm p_{eff}\geq 0$,
\item \textbf{WEC:} \quad$\rho_{eff}+p_{eff}\geq 0,
\quad\rho_{eff}\geq0$.
\end{itemize}
where $\rho_{eff}=\rho_m+\rho_c$ and $p_{eff}=p_m+p_c$. With the
help of Eqs.(\ref{6}) and (\ref{7}), we obtain
\begin{eqnarray}\nonumber
p_m&=&-\frac{\left(M'^2+2a'M'M-4Me^b\right)}{4M^2e^b\left(2+k^2\right)}
-\frac{f\left(\mathcal{G}\right)}{2\left(2+k^2\right)}+\frac{e^{-2b}}
{2+k^2}\left[\left(\frac{a'b'^2M'}{2M}\right.\right.\\\nonumber&+&\left.
\left.\frac{a'^2M'^2}{4M^2}-\frac{7a'b'M'^2}{M^2}+\frac{M'a'^3}{M}
+\frac{2a'a''M'}{M}+\frac{2a'M''M'}{M}-\frac{3a'M'^3}{2M^3}\right.
\right.\\\nonumber&+&\left.\left.\frac{11b'M'^3}{4M^3}+\frac{a''M'^2}{2
M^2}-\frac{a'^2e^b}{M}+\frac{a'b'e^b}{M}+\frac{4b'M'e^b}{M^2}
-\frac{2a''e^b}{M^2}-\frac{4M''e^b}{M^2}\right.\right.\\\nonumber&
+&\left.\left.\frac{4M'^2e^b}{M^3}+\frac{4M''^2}{M^2}-
\frac{4M''M'^2}{M^3}-\frac{M'^4}{2M^4}-\frac{a'b'^2M'}{2M}
-\frac{2b'M'^3}{M^2}+\frac{a'^4}{4}\right.\right.\\\nonumber&+
&\left.\left.\frac{4M'^3M''}{M^3}-\frac{a'^3b'}{2}
-a'b'a''+a''^2+\frac{a'^2b'^2}{4}+\frac{b'^2M'^2}{M^2}-\frac{4b'M'
M''}{M^2}\right.\right.\\\nonumber&+&\left.\left.\frac{a'M'M''}{M^2}
-\frac{4M''e^b}{M^2}-2M''M'^2\right)f_\mathcal{G}+\frac{a'^3}{2}
+a'^2M+a''a'M\right.\\\nonumber&+&\left.\left(\frac{2M'^3}{M^3}
-\frac{a'^2M'}{M}-\frac{5a'M'^2}{2M^2}+\frac{3b'M'^2}{2}+\frac{2a'e^b}{M}
+\frac{4M'e^b}{M^2}-\frac{2M'M''}{M^2}\right.\right.\\\label{9}&-&\left.
\left.\frac{2b'M''}{M}+\frac{4MM''}{M^2}-\frac{a'^3}{2}-a'a''\right)
f_\mathcal{G}'\right],
\\\nonumber
\rho_m&=&\frac{\left(-4M''M+2b'M'M+M'^2+4Me^b\right)}{4k^2e^bM^2}
+\frac{f_\mathcal{G}}{2k^2}-\frac{e^{-2b}}{k^2}\left(\frac{a'^4}{2}
-\frac{a'b'M'}{2M^2}\right.\\\nonumber&+&\left.\frac{M'a'b'^2}{M}
-\frac{M'^2a'b'}{M^2}-\frac{2M'a''b'}{M}+2a''^2-\frac{e^ba'^2}{M}
+\frac{e^ba'b'}{M}-\frac{3M'a'e^b}{M^2}\right.\\\nonumber&-&\left.\frac{2
e^ba''}{M}+\frac{2M''a'^2}{M}-\frac{2M''a'b'}{M}+\frac{3M''a'M'}{M^2}
+\frac{4M''a''}{M}+\frac{M'^2a'b'}{4M^2}+\frac{a'^4}{2}\right.
\\\nonumber&-&\left.\frac{a'M'^3}{2M^3}-\frac{a''M'^2}{2M^2}
-\frac{3a'^3b'}{4}-\frac{a'^2b'M'}{2M}+2a''a'^2-\frac{3a''a'b'}{2}
-\frac{a''M'^2}{M^2}\right.\\\nonumber&+&\left.\frac{b'^2a'^2}{8}
\right)f_\mathcal{G}+\frac{e^{-2b}}{k^2}\left(2b'a''-a'a''
+\frac{4M'a''}{M}-\frac{M'^2b'}{M^2}+\frac{4M''M'}{M^2}
-\frac{M'^3}{M^3}\right.\\\nonumber&+&\left.\frac{a'b'M'}{M}
-\frac{a'^3}{2}+\frac{3a'^2b'}{2}-\frac{a'^2M'}{M}
+\frac{2a'M'^2}{M^2}-a'b'\right)+\frac{e^{-b}}{k^2}
\left(\frac{2a'^2M'}{M}+a'^3\right.\\\nonumber&-&
\left.a'^2b'+2a'a''-e^{2a-2b}b'^2a'-\frac{2M'^2}{M^2}\right)
f_\mathcal{G}'+\left[\frac{4e^{-b}}{k^2M}+\frac{e^{-2b}}{k^2}
\left(\frac{M'^2}{M^2}-4a'b'\right.
\right.\\\label{10}&+&\left.\left.\frac{4M''}{M}
+3a'^2+5a''\right)\right]f_\mathcal{G}''.
\end{eqnarray}

For the traversability of WH, the basic property is the violation of
NEC in GR. This violation prevents the WH throat to shrink and leads
to the physically unrealistic WH solutions. The modified theories of
gravity provide $T_{\alpha\beta}^{eff}$ as an alternative source to
meet the violation of NEC. In this regard, these theories may have
an opportunity for usual matter configuration to fulfill the energy
constraints. Simplifying Eqs.(\ref{6}) and (\ref{7}), we obtain NEC
with respect to the effective stress-energy tensor as follows
\begin{equation}\label{11}
\rho_{eff}+p_{eff}=\frac{1}{2e^b}\left(\frac{M'a'}{M}
-\frac{2M''}{M}+\frac{M'^2}{M^2}+\frac{M'b'}{M}\right).
\end{equation}

\section{Point-like Lagrangian and Noether Symmetry Approach}

Here we use Lagrange multiplier technique to formulate the
Lagrangian for the action (\ref{1}). We take
\begin{equation}\label{12}
A=\int dr\sqrt{-g}[R+f(\mathcal{G})-\mu_1(\mathcal{G}
-\bar{\mathcal{G}})+\mathcal L_m],
\end{equation}
where $\sqrt{-g}=Me^\frac{a}{2}e^\frac{b}{2}$ and $\mathcal L_m=p_m$
while curvature scalar and GB invariant are
\begin{eqnarray*}
R&=&\frac{1}{e^b}\left(-\frac{a'^2}{2}+\frac{a'b'}{2}-\frac{a'M'}{M}
-\frac{2M''}{M}+\frac{b'M'}{M}+\frac{M'^2}{2M^2}-a''+\frac{2e^b}{M}\right)\\\nonumber
\bar{\mathcal{G}}&=&\frac{2e^{-2b}}{M^2}\left(a'^2M'^2-3b'a'M'^2-e^{b}a'^2+e^{b}a'b'+2a''M'^2
+4a'M'M''-2e^{b}a''\right).
\end{eqnarray*}
Varying the action (\ref{12}) relative to $\mathcal{G}$, we obtain
$\mu_1=f_\mathcal{G}(\mathcal{G})$ whereas the conservation of
energy-momentum tensor relative to perfect fluid gives
$p_{m}(r)=\rho_0e^{\frac{-a(1+w)}{2w}}$, ($\rho_o$ is integration
constant while $w$ denotes EoS parameter). Putting all
these values in Eq.(\ref{12}), it follows that
\begin{eqnarray}\nonumber
A&=&\int\left[Me^\frac{b+a}{2}\left\{R+f(\mathcal{G})
-\mathcal{G}f_\mathcal{G}+\rho_0e^{\frac{-a(1+w)}{2w}}+\frac{2e^{-2b}f_\mathcal{G}}{M^2}\left(a'^2M'^2
-3b'a'M'^2\right.\right.\right.\\\label{14}&-&\left.\left.\left.
e^{b}a'^2+e^{b}a'b'+2a''M'^2 +4a'M'M''-2e^{b}a''\right)
\right\}\right]dr.
\end{eqnarray}
In order to eliminate second order derivative, we integrate these
terms by parts and neglect boundary terms which leads to
\begin{eqnarray}\nonumber
\mathcal{L}\left(r,a,b,M,\mathcal{G},a',b',M',\mathcal{G}'\right)
&=&e^\frac{a+b}{2}M\left(R+f-\mathcal{G}f_\mathcal{G}+\rho_0e^{\frac{-a(1+w)}{2w}}\right)
+\frac{2e^\frac{a-3b}{2}}{M}\\\nonumber&\times&\left((4a'b'+2a'M')(M'^2-e^b)+2a'b'e^b\right)f_\mathcal{G}
\\\label{15}&-&\frac{4e^\frac{a-3b}{2}}{M}
\left(M'^2-e^b\right)a'\mathcal{G}'f_{\mathcal{G}\mathcal{G}}.
\end{eqnarray}

For static spherically symmetric metric, Hamiltonian of the
dynamical system and the Euler-Lagrange equation corresponding to
point-like Lagrangian are characterized as
\begin{eqnarray}\label{16}
\mathcal{H}=\sum_i q'_ip^i-\mathcal{L},\quad\frac{\partial
\mathcal{L}}{\partial q_i}-\frac{d}{dr}\left(\frac{\partial
\mathcal{L}}{\partial q'_{i}}\right)=0,
\end{eqnarray}
where $q^i$ are generalized coordinates. The differential of
Lagrangian with respect to the configuration space
$(a,b,M,\mathcal{G})$ gives
\begin{eqnarray}\nonumber
&&\frac{e^\frac{a}{2}e^\frac{b}{2}}{2}\left(-\mathcal{G}
Mf_\mathcal{G}+RM+Mf_\mathcal{G}-Mp\right)+\frac{1}{2}
\left(4a'b'e^\frac{a-b}{2}+\frac{a'M'^3}{M^2}e^\frac{a-3b}{2}
\right.\\\nonumber&-&\left.\frac{a'^2M'^2}{M}e^\frac{a-3b}{2}
+\frac{3a'b'M'^2}{M}e^\frac{a-3b}{2}\right)
f_\mathcal{G}+\frac{1}{2}f_{\mathcal{G}\mathcal{G}}
\left(\frac{-3\mathcal{G}'a'M'^2}{M}e^\frac{a-3b}{2}
-4a'\mathcal{G}'\right.\\\nonumber&\times&\left.
e^\frac{a-b}{2}\right)-\frac{1}{2}a'f_\mathcal{G}
\left(4b'e^\frac{a-b}{2}+\frac{M'^3e^\frac{a
-3b}{2}}{M^2}-\frac{2a'M'^2e^\frac{a-3b}{2}}{M}
+\frac{3b'M'^2e^\frac{a-3b}{2}}{M}\right)\\
\nonumber&-&\frac{1}{2}b'\left(4b'e^\frac{a-b}{2}
+\frac{M'^3e^\frac{a-3b}{2}}{M^2}-\frac{2a'
M'^2e^\frac{a-3b}{2}}{M}+\frac{3b'M'^2
e^\frac{a-3b}{2}}{M}\right)f_\mathcal{G}
-\left(-4b'^2\right.\\\nonumber&\times&
\left.e^\frac{a-b}{2}-\frac{2M'^4e^\frac{a-3b}{2}}{M^3}
-\frac{2M'^3b'e^\frac{a-3b}{2}}{M^2}+\frac{3M'^2M''
e^\frac{a-3b}{2}}{M^2}-\frac{2a''M'^2e^\frac{a-3b}{2}}{M}
+4\right.\\\nonumber&\times&\left.b''e^\frac{a-b}{2}
+\frac{2a'M'^3e^\frac{a-3b}{2}}{M^2}-\frac{4a'M'M''
e^\frac{a-3b}{2}}{M}+\frac{4a'b'M'^2e^\frac{a-3b}{2}}{M}
+\frac{3b''M'^2e^\frac{a-3b}{2}}{M}\right.\\\nonumber&+&\left.
\frac{6b'M''M'e^\frac{a-3b}{2}}{M}-\frac{6M'^2b'^2e^\frac{a
-3b}{2}}{M}-\frac{3M'^3b'e^\frac{a-3b}{2}}{M^2}\right)
f_\mathcal{G}-\frac{1}{2}a'\left(-3\mathcal{G}'
e^\frac{a-b}{2}\right.\\\nonumber&-&\left.
\frac{4\mathcal{G}'M'^2e^\frac{a-3b}{2}}{M}\right)
f_{\mathcal{G}\mathcal{G}}-\frac{1}{2}b'\left
(\frac{-3\mathcal{G}'M'^2e^\frac{a-3b}{2}}{M}
-4\mathcal{G}'e^\frac{a-b}{2}\right)f_{\mathcal{G}
\mathcal{G}}-\mathcal{G}'\left(-4\mathcal{G}'\right.
\\\nonumber&\times&\left.e^\frac{a-b}{2}-
\frac{3\mathcal{G}'M'^2e^\frac{a-3b}{2}}{M}\right)
f_{\mathcal{G}\mathcal{G}\mathcal{G}}-\left
(\frac{-3\mathcal{G}''M'^2e^\frac{a-3b}{2}}
{M}-\frac{6\mathcal{G}'M'M''e^\frac{a-3b}{2}}{M}
-4\right.\\\label{17}&\times&\left.\mathcal{G}''e^\frac{a
-b}{2}+\frac{3\mathcal{G}'M'^3e^\frac{a-3b}{2}}{M^2}
+\frac{6\mathcal{G}'b'M'^2e^\frac{a-3b}{2}}{M}
+4\mathcal{G}'b'e^\frac{a-b}{2}\right)f_{\mathcal{G}\mathcal{G}}=0,
\\\nonumber
&&\frac{1}{2M^4}\left[-\left(12a'M'M''M^3
e^\frac{a-3b}{2}+9a'b'M'^2M^3e^\frac{a-3b}{2}
-9a'b'M'^2M^3e^\frac{a-3b}{2}\right.\right.
\\\nonumber&-&\left.\left.M^5e^\frac{a+b}{2}
+4a'^2M^4e^\frac{a-b}{2}+8a''M^4e^\frac{a
-b}{2}+\mathcal{G}M^5e^\frac{a+b}{2}+
3a'M'^3M^2e^\frac{a-3b}{2}\right.\right.
\\\nonumber&+&\left.\left.pM^5e^\frac{a+b}{2}
-RM^5e^\frac{a+b}{2}-6a'M'^3M^2e^\frac{a-3b}{2}
+6a''M'^2M^3e^\frac{a-3b}{2}\right)f_\mathcal{G}
+\left(\right.\right.\\\label{18}&+&\left.\left.
4a'\mathcal{G}'M^4e^\frac{a-b}{2}-9a'M'^2
M^3e^\frac{a-3b}{2}+6a'\mathcal{G}' M'^2M^3e^\frac{a-3b}{2}\right)
f_{\mathcal{G}\mathcal{G}}\right]=0,
\\\nonumber
&&e^\frac{a}{2}e^\frac{b}{2}\left(
-\mathcal{G}f_\mathcal{G}+R+f_\mathcal{G}
-p\right)+e^\frac{a-3b}{2}\left[\left(
\frac{4M'^3a'}{M^3}-\frac{5M'^2a'^2}{2
M^2}+\frac{15M'^2a'b'}{2M^2}\right.\right.
\\\nonumber&+&\left.\left.\frac{a'^3
M'}{M}-\frac{3M'a'b'}{M}-\frac{3M'a'^2b'}{M}
-\frac{9M'a'b'^2}{M}-\frac{3M'^2a''}{M^2}
-\frac{6M''M'a'}{M^2}\right.\right.
\\\nonumber&+&\left.\left.\frac{4M'a''a'}{M}
+\frac{2M''a'^2}{M}-\frac{6M'b'a''}{M}
-\frac{6M'b''a'}{M}-\frac{6M''b'a'}{M}\right)
f_\mathcal{G}+\left(\frac{3\mathcal{G}'a'^2M'}{M}
\right.\right.\\\nonumber&-&\left.\left.
\frac{9a'\mathcal{G}'b'M'}{M}+\frac{6a'\mathcal{G}''M'
}{M}+\frac{6a''\mathcal{G}'M'}{M}
+\frac{6a'\mathcal{G}'M''}{M}-\frac{6a'M'^2\mathcal{G}'}{M^2}\right)
f_{\mathcal{G}\mathcal{G}}\right.\\\label{19}&+
&\left.\frac{6\mathcal{G}'^2a'M'}{M}f_{\mathcal{G}
\mathcal{G}\mathcal{G}}\right]=0,
\\\nonumber
&&\frac{1}{2M^2}\left[-2M^3e^\frac{a}{2}e^\frac{b}{2}f_\mathcal{G}-
\left(-3a'^2M'^2Me^\frac{a-3b}{2}-4a'^2M^2e^\frac{a-b}{2}+6a'M'^3e^\frac{a
-3b}{2}\right.\right.\\\nonumber&+&\left.\left.4a'b'M^2e^\frac{a-b}{2}
+9a'b'M'^3e^\frac{a-3b}{2}-12a'M'M''Me^\frac{a-3b}{2}-6a''M'^2Me^\frac{a
-3b}{2}\right.\right.\\\label{20}&-&\left.\left.8a''M^2e^\frac{a-b}{2}\right)
f_{\mathcal{G}\mathcal{G}}+\left(6a'M'^2\mathcal{G}'Me^\frac{a
-3b}{2}+8a'\mathcal{G}'M^2e^\frac{a-b}{2}\right)f_{\mathcal{G}\mathcal{G}
\mathcal{G}}\right]=0.
\end{eqnarray}

In order to solve the system of non-linear differential equations,
Noether symmetry is recognized as a significant tool. The
physical properties of any dynamical structure can be illustrated by
the respective Lagrangian which narrates the energy density as well
as the presence of symmetries of the system. Noether theorem can be
stated as a group generator that provides conserved quantity only if
point-like Lagrangian shows constant behavior under a continuous
group. To analyze the associated conserved quantity as well as the
existence of Noether symmetry for the spherical system, we take a
vector field $K$ \cite{25,25a}
\begin{equation}\label{21}
K=\tau\left(r,q^i\right)\frac{\partial}{\partial
r}+\zeta^i\left(r,q^i\right)\frac{\partial}{\partial q^i},
\end{equation}
where $\tau$ and $\zeta^i$ are unknown coefficients while $r$ acts
as an affine parameter of $K$. This leads to uniqueness of the
vector field in the tangent space.

The corresponding invariance condition is characterized by
\begin{equation}\label{22}
K^{[1]}\mathcal{L}+(D\tau)\mathcal{L}=D B(r,q^i).
\end{equation}
Here $B$ signifies the boundary term, $K^{[1]}$ and $D$ represent
the first order expansion and total derivative, respectively given
by
\begin{equation}\label{23}
K^{[1]}=K+\left(D\zeta^i-q'^iD\tau\right)\frac{\partial}{\partial
q'^i},\quad D=q'^i\frac{\partial}{\partial
q^i}+\frac{\partial}{\partial r}.
\end{equation}
Invariance condition (\ref{22}) leads to the Noether symmetries
which represent the related conserved parameters in terms of first
integral. Under translation with respect to time as well as
position, if the Lagrangian shows constant behavior, then the first
integral describes conservation of energy as well as the linear
momentum whereas rotationally symmetric Lagrangian provides angular
momentum conservation \cite{26}. The first integral for invariance
condition (\ref{22}) is expressed in the form
\begin{equation}\label{24}
\Sigma=B-\tau
\mathcal{L}-\left(\zeta^i-q'^i\tau\right)\frac{\partial
\mathcal{L}}{\partial q'^i}.
\end{equation}

The vector field and first order expansion for the configuration
space become
\begin{eqnarray}\nonumber
K&=&\tau\frac{\partial}{\partial r}+\alpha\frac{\partial}{\partial
a}+\beta\frac{\partial}{\partial b}+\gamma\frac{\partial}{\partial
M}+\delta\frac{\partial}{\partial \mathcal{G}}, \quad
K^{[1]}=\tau\frac{\partial}{\partial
r}+\alpha\frac{\partial}{\partial a}+\beta\frac{\partial}{\partial
b}\\\label{25}&+&\gamma\frac{\partial}{\partial
M}+\delta\frac{\partial}{\partial
\mathcal{G}}+\alpha'\frac{\partial}{\partial
a'}+\beta'\frac{\partial}{\partial
b'}+\gamma'\frac{\partial}{\partial
M'}+\delta'\frac{\partial}{\partial \mathcal{G}'},
\end{eqnarray}
where the unknown parameters of vector field having the radial
derivative are given by
\begin{equation}\label{26}
\sigma_j'=D\sigma_j-q'^iD\tau,\quad j=1,...,4,
\end{equation}
where $\sigma_{j} (j=1,2,3,4)$ denote $\alpha$, $\beta$, $\gamma$
and $\delta$, respectively. Comparing the coefficients of $a'^2b'$,
$a'b'^2M'^2$, $a'b'M'^3$ and $M'^2\mathcal{G}'^2a'$, we obtain
\begin{eqnarray}\label{27}
&&\tau,_af_\mathcal{G}=0,\quad\tau ,_bf_\mathcal{G}=0,\quad\tau
,_Mf_\mathcal{G}=0 ,\quad\tau
,_\mathcal{G}f_{\mathcal{G}\mathcal{G}}=0,
\end{eqnarray}
This leads to a trivial solution for $f_\mathcal{G}=0$. For
non-trivial solution, we assume $f_\mathcal{G}\neq0$ and compare the
coefficients of $a'$, $b'$, $M'$, $\mathcal{G}'$, $b'^2$, $M'^2$,
$\mathcal{G}'^2$, $\mathcal{G}'M'^2$, $a'b'M'$,
$a'^2M'\mathcal{G}'$, $a'M'b'^2$ and $a'M'\mathcal{G}'^2$ leading to
following equations
\begin{eqnarray}\label{27a}
&&4e^{a-b/2}[f_\mathcal{G}(-\beta_{,r}-\gamma_{,r})+\delta_{,r}f_\mathcal{GG}]=MB_{,a},
\quad -4e^{a-b/2}\alpha_{,r}f_\mathcal{G}=MB_{,b},\\\label{27a1}&&
-4e^{a-b/2}\alpha_{,r}f_\mathcal{G}=MB_{,M},\quad
4e^{a-b/2}\alpha_{,r}f_{\mathcal{G}\mathcal{G}}=MB_{,\mathcal{G}},
\\\label{27a2}&&\alpha,_bf_\mathcal{G}=0,\quad\alpha,_Mf_\mathcal{G}=0,
\quad\alpha,_{\mathcal{G}}f_{\mathcal{G}\mathcal{G}}=0,\quad\alpha,_{r}f_{\mathcal{G}
\mathcal{G}}=0,\\\label{27a3}&&4\gamma,_rf_\mathcal{G}=0,\quad4\gamma,_af_{\mathcal{G}\mathcal{G}}=0
,\quad8\gamma,_bf_\mathcal{G}=0,\quad4\gamma,_{\mathcal{G}}f_{\mathcal{G}\mathcal{G}}=0.
\end{eqnarray}
For $f_{\mathcal{G}}\neq0$, we compare remaining coefficients and
obtain over determined system of equations as follows
\begin{eqnarray}\label{28}
&&\tau,_a=0,\quad
\tau,_b=0,\quad\tau,_M=0,\quad\tau,_{\mathcal{G}}=0,\quad\gamma,_a=0,\quad\gamma,_{\mathcal{G}}=0\\\label{28a}&&
B,_b=0,\quad B,_M=0,\quad
B,_{\mathcal{G}}=0,\quad\gamma,_r=0,\quad\gamma,_b=0\\\label{28b}&&
\alpha,_r=0,\quad\alpha,_b=0,\quad\alpha,_M=0,\quad\alpha,_{\mathcal{G}}=0,
\\\label{29}&&4e^{a-b/2}[-f_\mathcal{G}\beta_{,r}+\delta_{,r}f_\mathcal{GG}]=MB_{,a},
\\\label{30}&&\beta_{,r}f_\mathcal{G}-\delta_{,r}f_\mathcal{GG}=0,\quad\beta_{,a}f_\mathcal{G}
-\delta_{,a}f_\mathcal{GG}=0,\\\label{31}&&(-\alpha+\beta+M^{-1}\gamma-\delta-\alpha,_{a}
-\beta,_{b}+\tau,_{r})f_\mathcal{G}+\delta_{,b}f_\mathcal{GG}=0,\\\label{32}&&(-\alpha+\beta+M^{-1}\gamma-\alpha,_{a}
-\beta,_{M}-\gamma,_{M}+\tau,_{r})f_\mathcal{G}-(\delta-\delta_{,M})f_\mathcal{GG}=0,
\\\label{33}&&2(\alpha-3\beta-M^{-1}\gamma+\alpha,_{a}
+\beta,_{b}+2\gamma,_{M}-2\tau,_{r})f_\mathcal{G}+(2\delta-\delta_{,b})f_\mathcal{GG}=0,
\\\label{34}&&(\alpha-3\beta-M^{-1}\gamma+\alpha,_{a}
+2\beta,_{M}+3\gamma,_{M}-3\tau,_{r})f_\mathcal{G}+(\delta-\delta_{,M})f_\mathcal{GG}=0,
\\\label{35}&&-\beta,_{\mathcal{G}}f_\mathcal{G}+f_{\mathcal{G}\mathcal{G}}(\alpha-\beta-M^{-1}\gamma
+\alpha,_{a}+\delta,_{\mathcal{G}}-\tau,_{r})+\delta
f_{\mathcal{G}\mathcal{G}\mathcal{G}}=0,\\\nonumber&&2\beta,_{\mathcal{G}}f_\mathcal{G}
+f_{\mathcal{G}\mathcal{G}}(\alpha+3\beta+M^{-1}\gamma
-\alpha,_{a}-2\gamma,_{M}-\delta,_{\mathcal{G}}+2\tau,_{r})+2\delta
f_{\mathcal{G}\mathcal{G}\mathcal{G}}=0,\\\label{36}\\\nonumber&&
e^{a+b/2}M\left[(R+f-\mathcal{G}f_{\mathcal{G}})\left(\frac{\alpha}{2}+\frac{\beta}{2}
+\frac{\gamma}{M}+\tau,_{r}\right)+\rho_0e^{\frac{-a(1+w)}{2w}}\left(\frac{\alpha}{2w}+\frac{\beta}{2}
\right.\right.\\\label{37}&+&\left.\left.\frac{\gamma}{M}+\tau,_{r}\right)-\delta\mathcal{G}
f_{\mathcal{G}\mathcal{G}}\right]=B,_{r}.
\end{eqnarray}

Here we solve Eqs.(\ref{28})-(\ref{37}) for three different choices
of parameters given by
\begin{itemize}
\item $\beta(r,a,b,M,\mathcal{G})=0,\quad\delta(r,a,b,M,\mathcal{G})=0$,
\item
$\beta(r,a,b,M,\mathcal{G})=0,\quad\delta(r,a,b,M,\mathcal{G})\neq0$
or vice versa.
\item $\beta(r,a,b,M,\mathcal{G})\neq0,\quad\delta(r,a,b,M,\mathcal{G})\neq0$.
\end{itemize}

\section{$f(\mathcal{G})$ Models and Wormhole solutions}

In order to evaluate unknown parameters of symmetry generators and
explicit solution of $f$, we consider above mentioned possibilities
of $\beta$ and $\delta$.

\subsection*{Case I: $\beta=\delta=0$}

In this case, we obtain
\begin{eqnarray}\nonumber
\alpha&=&\xi_1+\xi_2e^{-a},\quad\gamma=M(\xi_3-\xi_4),\quad\tau=\xi_5r+\xi_6,
\\\label{38}f(\mathcal{G})&=&\xi_1\mathcal{G}+\xi_2,\quad B(r,a),_a=0,
\end{eqnarray}
where $\xi_{i's}$ denotes integration constants and explicit form of
$f$ corresponds to linear model which is compatible with
Gauss-Bonnet gravity. The coefficient of boundary term, symmetry
generator and $f(\mathcal{G})$ solution satisfy
Eqs.(\ref{28})-(\ref{36}). Consequently, the symmetry generator and
the first integral yield
\begin{eqnarray}\nonumber
K&=&\xi_1\frac{\partial}{\partial r}+\xi_2(\frac{\partial}{\partial
a}+\frac{\partial}{\partial b}),
\\\nonumber
\Sigma&=&\xi_3+\xi_4r^3-\xi_0\left[e^\frac{a}{2}e^\frac{b}{2}\left(r^2
R+r^2f_\mathcal{G}-r^2\mathcal{G}f_\mathcal{G}-r^2p\right)+e^\frac{a}{2}
e^\frac{b}{2}\left(\frac{4a'b'}{e^b}\right.\right.\\\nonumber&+&\left.\left.
\frac{8a'}{re^{2b}}-\frac{4a'^2}{e^{2b}}+\frac{12a'b'}{e^{2b}}\right)
f_\mathcal{G}+\left(\frac{-12a'\mathcal{G}'}{e^{2b}}-\frac{4a'
\mathcal{G}'}{e^b}\right)f_{\mathcal{G}\mathcal{G}}\right]-\xi_2
e^\frac{a}{2}e^\frac{b}{2}\\\nonumber&\times&\left[\left(\frac{4b'}{e^b}
+\frac{8}{re^{2b}}-\frac{8a'}{e^{2b}}+\frac{12b'}{e^{2b}}\right)
f_\mathcal{G}+\left(\frac{-12\mathcal{G}'}{e^{2b}}-\frac{4
\mathcal{G}'}{e^b}\right)f_{\mathcal{G}\mathcal{G}}\right].
\end{eqnarray}
We insert symmetry generators, $f(\mathcal{G})$ model in
Eq.(\ref{37}) with $B,_{r}=\xi_7$, $M=r^2$,
$e^b=\left(1-h(r)/r\right)^{-1}$ and $a(r)=-k/r$ (where $k$ is
positive constant) leading to
\begin{eqnarray}\nonumber
&&\frac{1}{2 w}\left(-\sqrt{\frac{r}{r-h(r)}} e^{\frac{k (2 w+1)}{2
r w}} \xi_2\rho_0r^2+2\rho_0\sqrt{\frac{r}{r-h(r)}} r^2
\left(w-\frac{1}{2}\right)\xi_3e^{\frac{k}{2 r
w}}+3\right.\\\nonumber&\times&\left.w\left( \left(\frac{r^2 (k+8
r)h'(r)+\left(k^2+3 k r+24 r^2\right)h(r)-k^2 r-4 k r^2-32 r^3+4
r}{2
r^3}\right.\right.\right.\\\label{39}&+&\left.\left.\left.r^2
\xi_1\right)\left(e^{-\frac{k}{2
r}} \xi_3+\frac{e^{\frac{k}{2 r}}\xi_2}{3}\right)
\sqrt{\frac{r}{r-h(r)}}-\frac{2\xi_7}{3}\right)\right)=0.
\end{eqnarray}
In order to study the geometry, traversability and physical
viability of WH in the presence of phantom energy, we consider
$w=-1$ and solve this non-linear equation numerically to construct
graphical analysis of the shape function. This analysis leads to
measure compatibility of linear $f(\mathcal{G})$ model with viable
models under the condition of regular and positive derivatives of
$f(\mathcal{G})$ function \cite{28}. Furthermore, we explore the
possibility of traversable WHs through graphical interpretation of
effective NEC. The graphical analysis of energy bounds, i.e., NEC,
WEC, SEC and DEC help to explore the presence/absence of ordinary
matter.
\begin{figure}\center
\epsfig{file=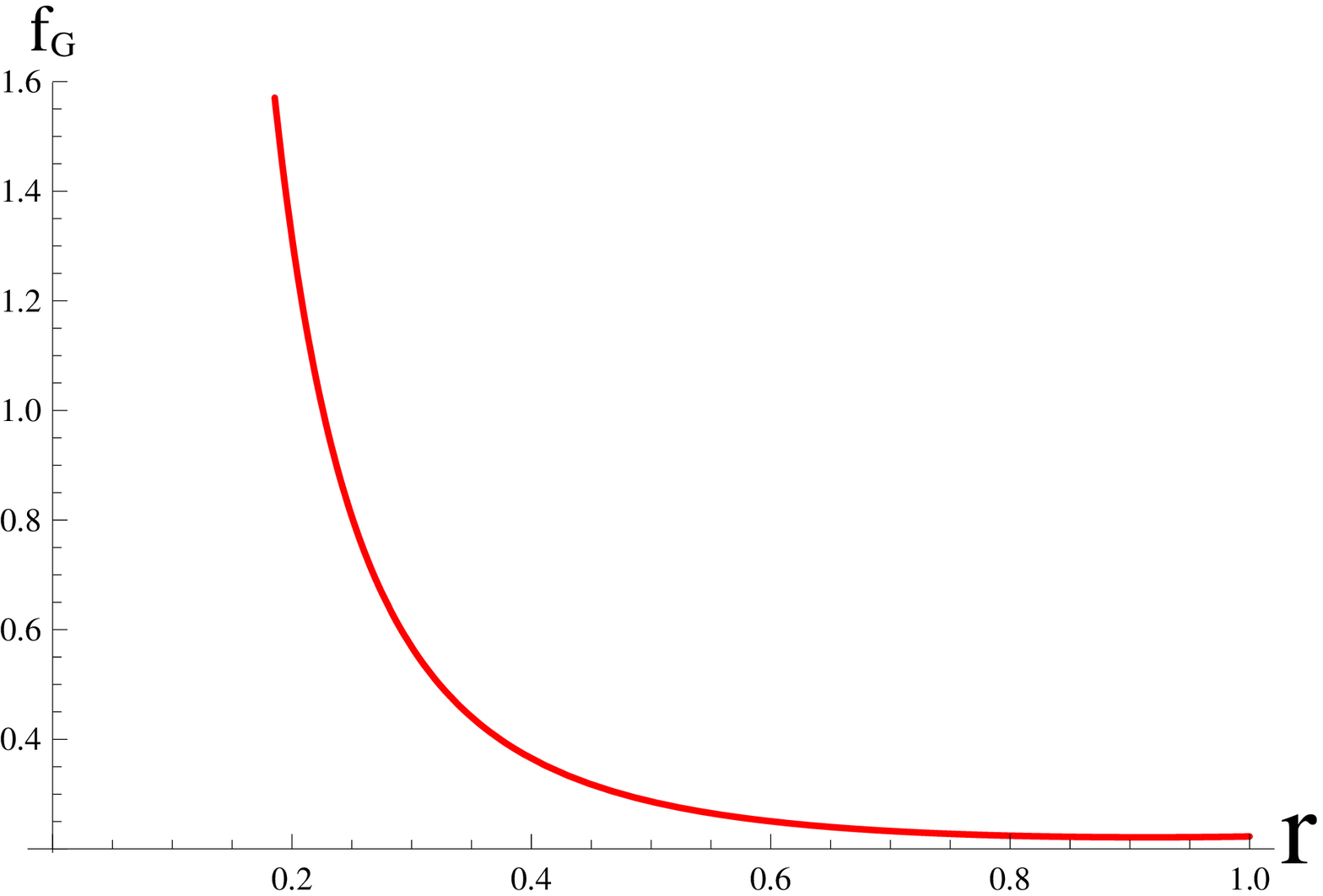,width=0.45\linewidth}
\epsfig{file=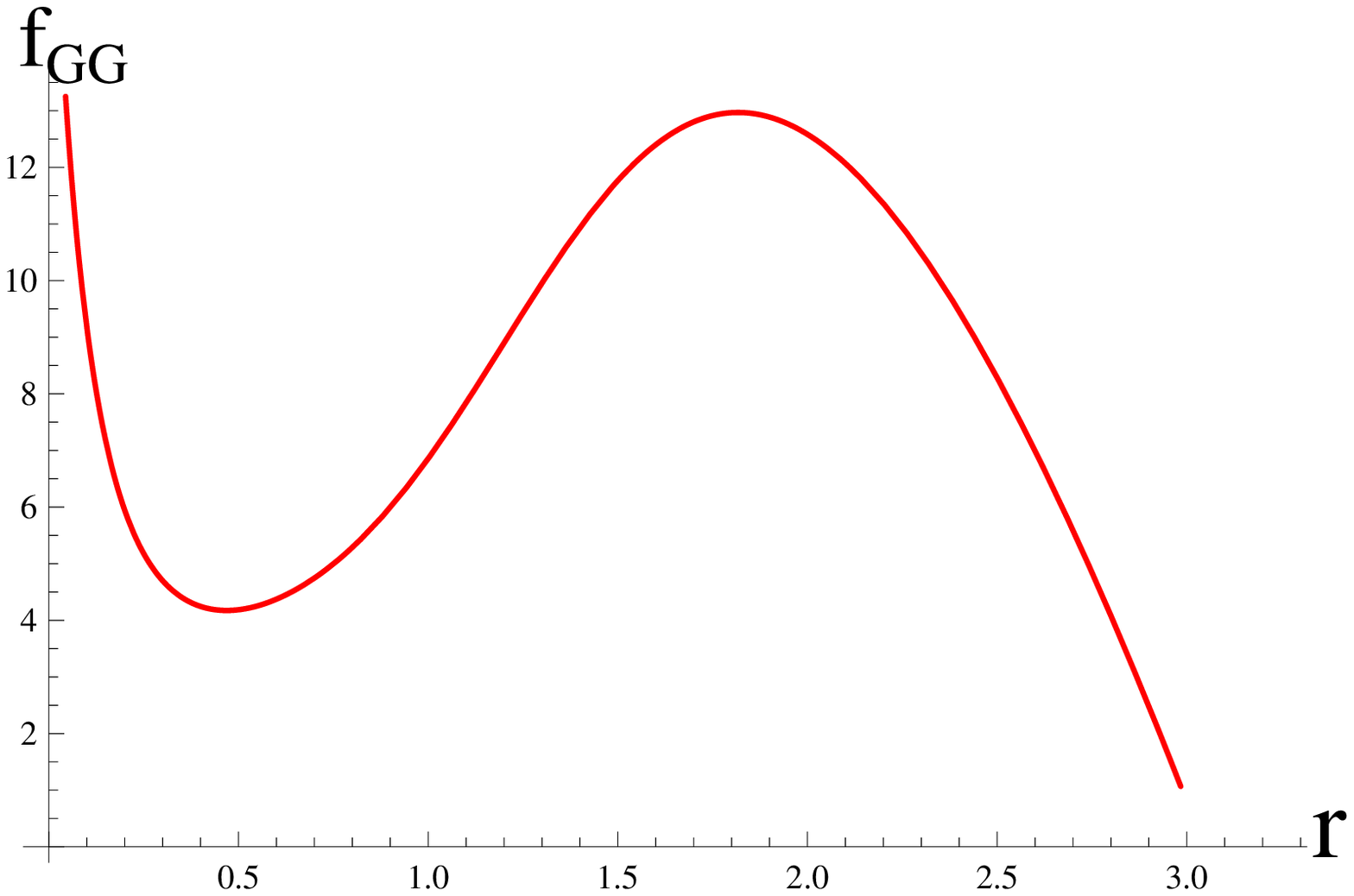,width=0.45\linewidth} \caption{Evolution of
$f_{\mathcal{G}}$ and $f_{\mathcal{G}{\mathcal{G}}}$ versus $r$ for
$k=0.005$, $\xi_1=0.01$, $\xi_2=0.85$, $\xi_3=0.1$, $\xi_7=0.5$,
$w=-1$ and $\rho_0=0.5$.}
\end{figure}
\begin{figure}\center
\epsfig{file=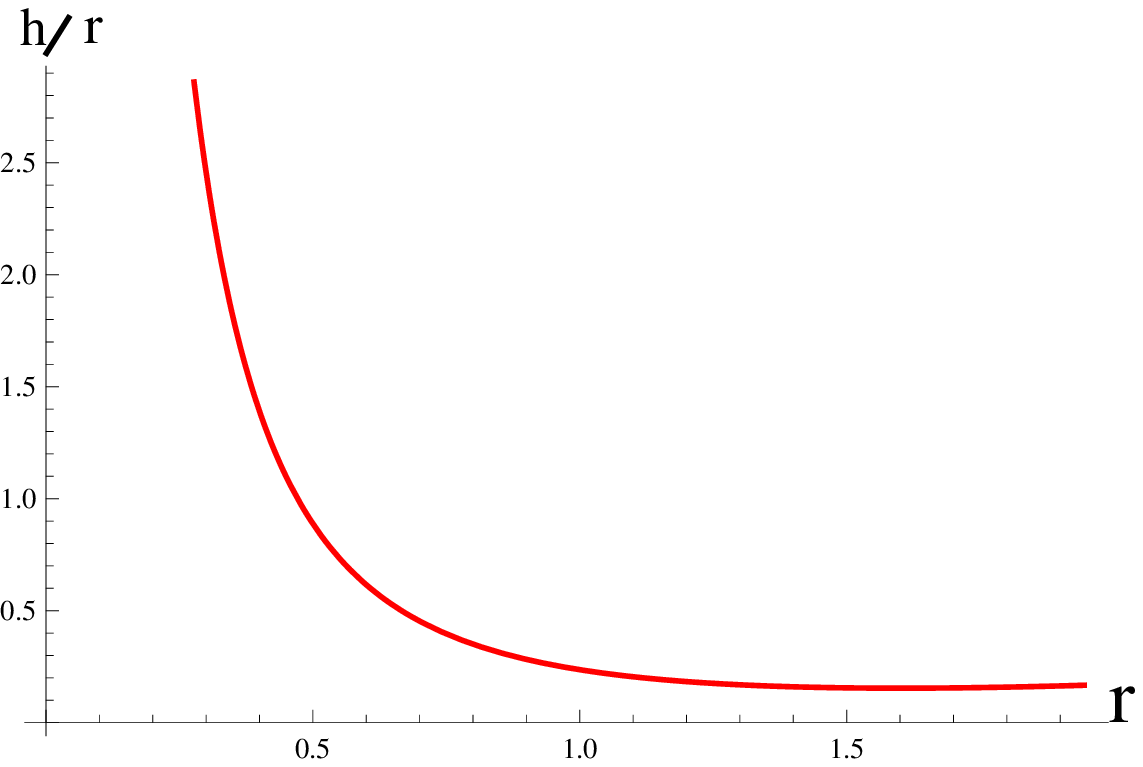,width=0.45\linewidth}
\epsfig{file=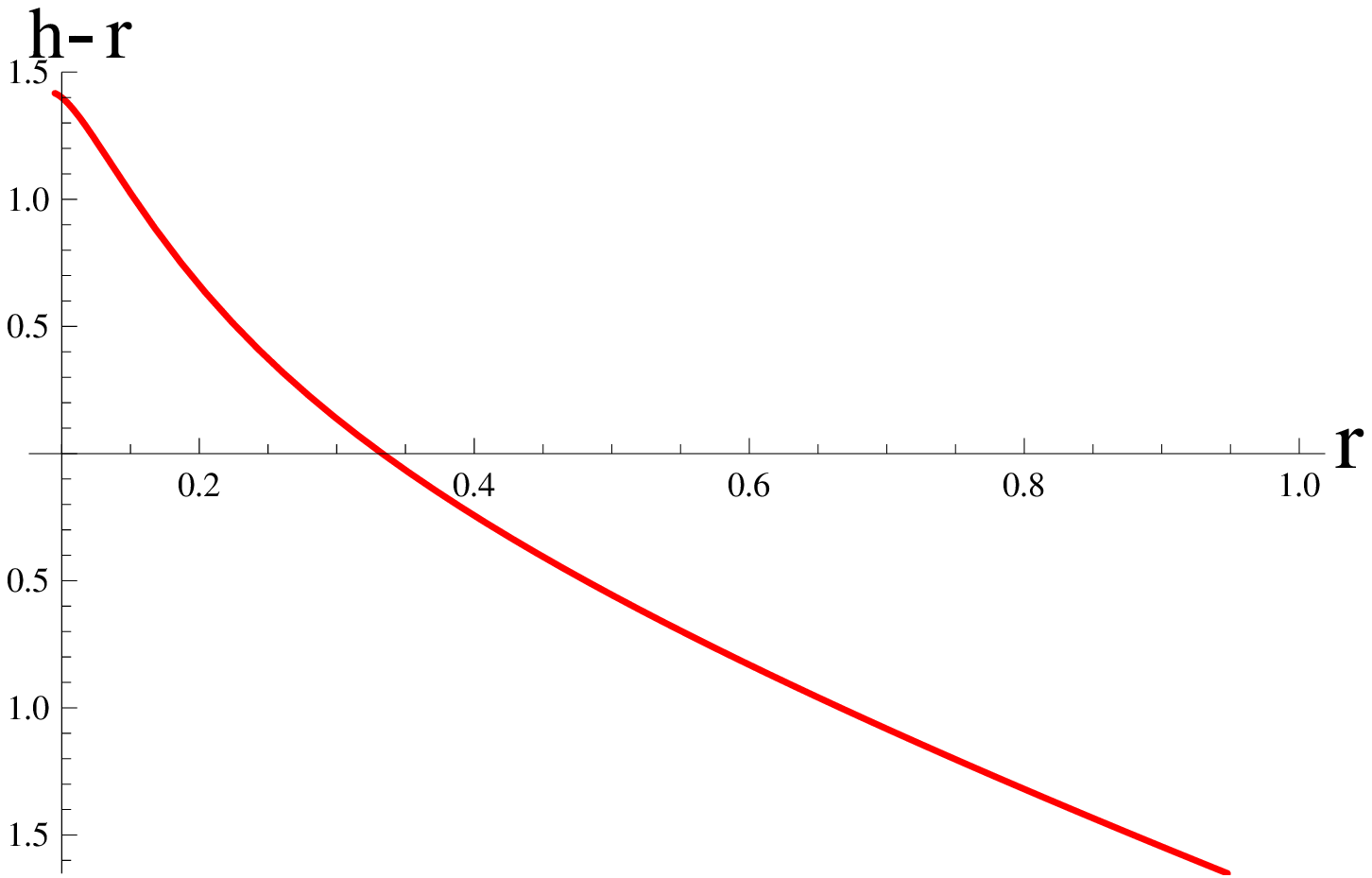,width=0.45\linewidth}\caption{Variation of the
shape function versus $r$.}
\end{figure}

In both plots of Figure \textbf{1}, the positively evolving curves
represent that $f(\mathcal{G}$) satisfies viability constraints as
$f_{\mathcal{G}}>0$ and $f_{\mathcal{G}\mathcal{G}}>0$. For linear
modified GB function, the WH geometry is analyzed in the context of
accelerated expansions ($w=-1$) in figure \textbf{2}. The left plot
indicates WH geometry to be asymptotically flat in a very short
interval of $r$ as $h/r\rightarrow0$ as $r\rightarrow1$. In the
right plot, the trajectory identifies throat of WH at $r_0=0.34$ and
the derivative of the shape function at this point remains positive,
i.e., $\frac{dh(r_0)}{dr}<1$. In the presence of accelerated
expansion of cosmos, the graphical analysis of WH geometry shows
that the configuration is compatible with Morris-Thorne WH proposal.
\begin{figure}\center
\epsfig{file=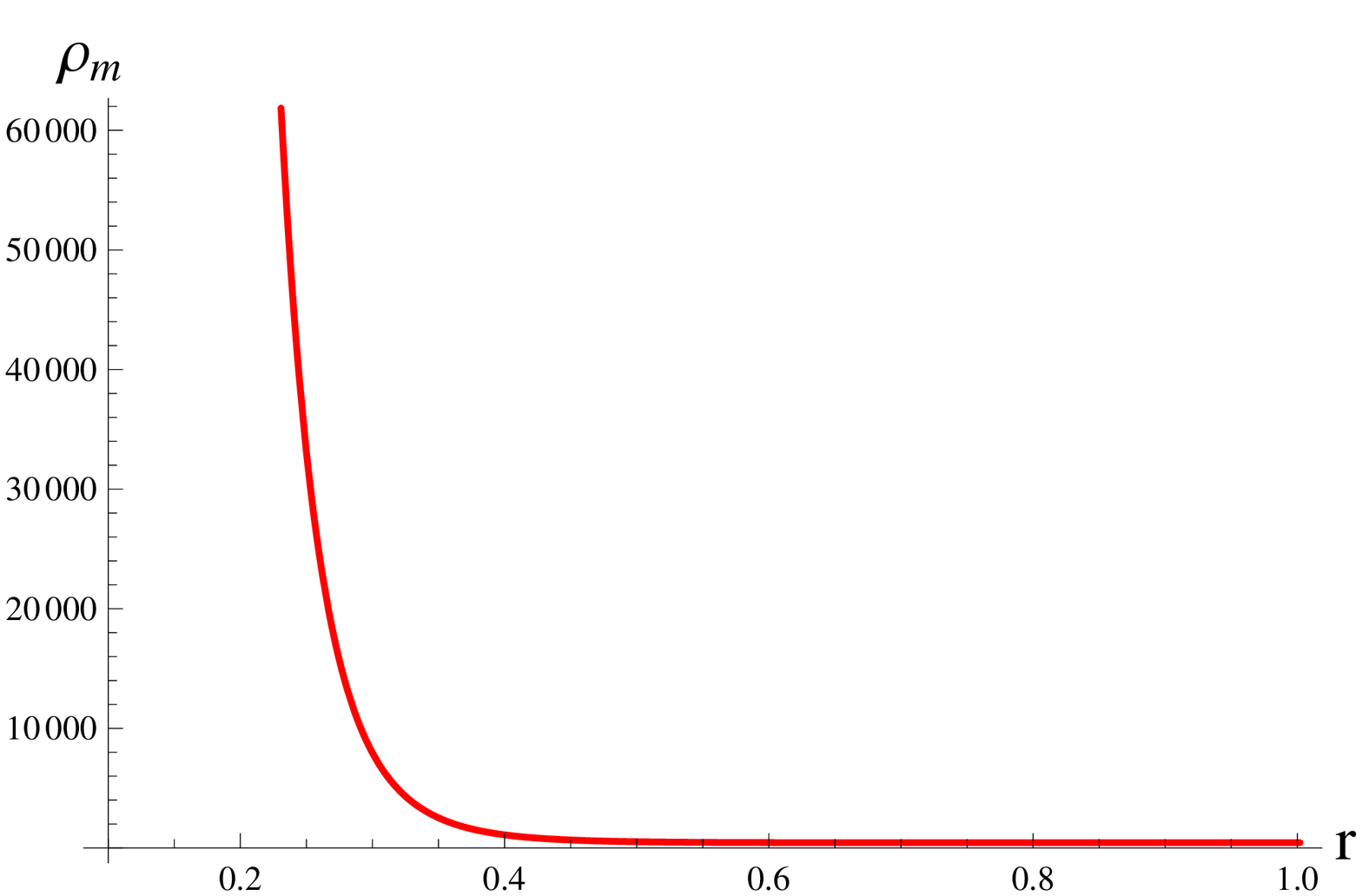,width=0.5\linewidth}\epsfig{file=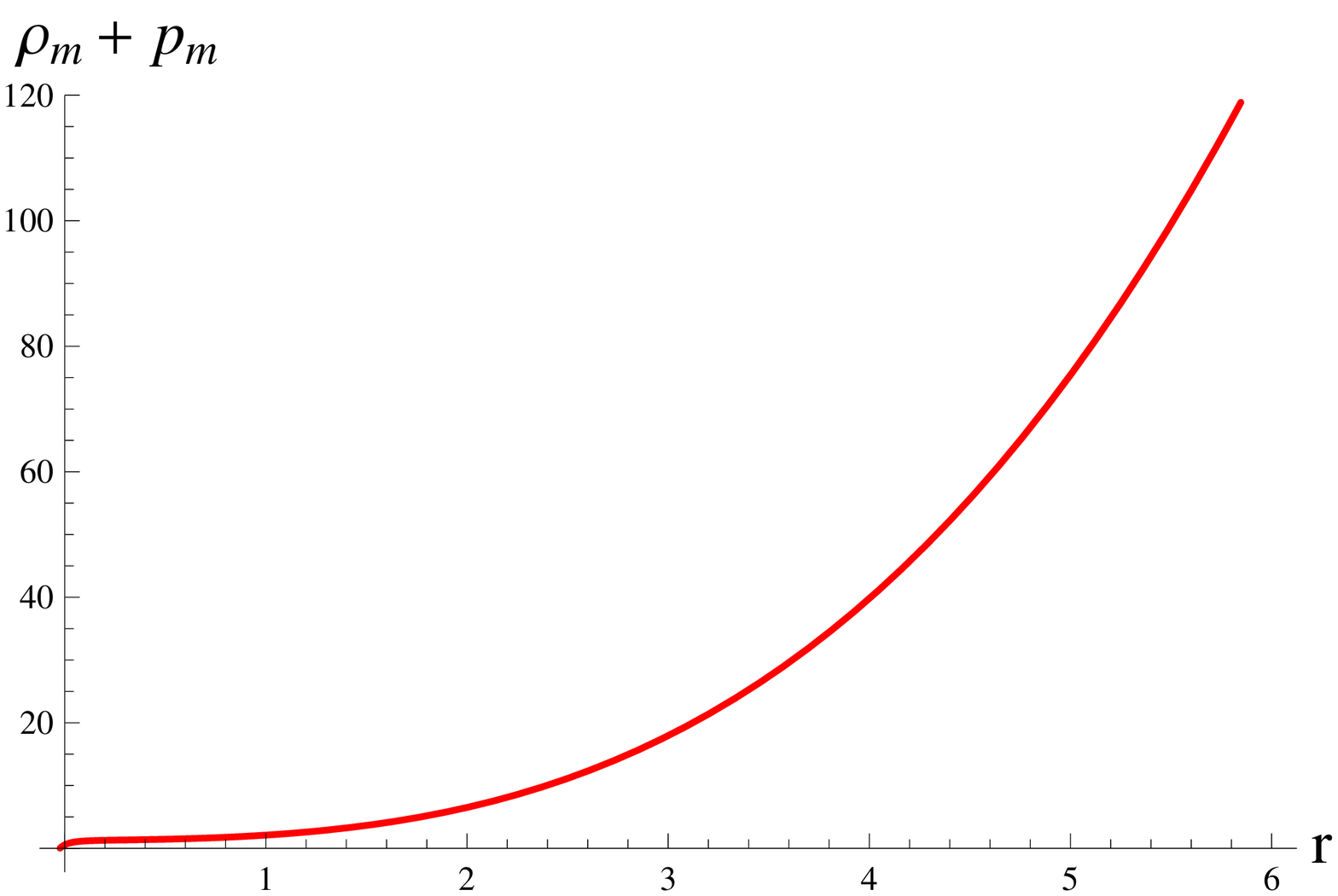,width=0.5\linewidth}
\caption{Evolution of energy bounds versus $r$ for $w=-1$.}
\end{figure}
\begin{figure}\center
\epsfig{file=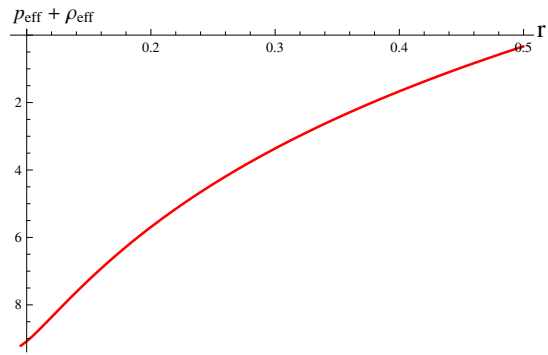,width=0.55\linewidth} \caption{Evolution of
effective NEC versus $r$ for $w=-1$.}
\end{figure}

The WH configuration is more significant if it is supported by
ordinary matter, i.e., the normal matter that satisfies energy
bounds. The criteria of energy bounds indicates that NEC is the
weakest condition as the violation of NEC leads to inconsistent
behavior of WEC, SEC and DEC. If a matter distribution follows DEC
then WEC and NEC holds trivially while SEC needs to be checked
separately. In order to examine realistic nature of WH, we discuss
the evolution of energy density and pressure of normal matter in
figure \textbf{3}. The graphical interpretation indicates that
$\rho_m\geq0$ and $\rho_m+p_m\geq0$. This behavior of matter
variables indicates that the WH is physically viable inside the
throat ($r_0=0.34$). To study traversable behavior of WH, we
substitute $a(r)=-k/r$ and Eq.(\ref{39}) in (\ref{11}) leading to
\begin{equation}\nonumber
p_{eff}+\rho_{eff}=-\frac{h(r)}{r^3}+\frac{h'(r)}{r^2}
+\frac{k(r-h(r))}{r^4}.
\end{equation}
For traversable WH, the violation of effective NEC
$(\rho_{eff}+p_{eff}<0)$ is required which also fulfills the
flaring-out condition. Figure \textbf{4} shows negatively increasing
curve which indicates that $p_{eff}+\rho_{eff}<0$ implying existence
of traversable WH solution. For linear $f(\mathcal{G})$ model, the
WH is found to be traversable as well as physically viable in the
presence of accelerating phases of cosmos.

\subsection*{Case II: $\beta=0,\quad\delta\neq0$}

For $\beta=0$, we solve the Eqs.(\ref{30})-(\ref{36}) and obtain
\begin{eqnarray}\nonumber
\alpha&=&\chi_1+\chi_2e^{-a},\quad\gamma=M(\chi_3Y(M)-\chi_4),
\quad\tau=\chi_5r+\chi_6,
\\\label{38a}f(\mathcal{G})&=&\chi_7\mathcal{G}^2+\chi_8\mathcal{G}
+\xi_1,\quad\delta=\chi_3Y(M),
\end{eqnarray}
where $\chi_{i's}$ are constants of integration while the explicit
form of $f$ corresponds to quadratic model. Solving Eq.(\ref{29})
for above solutions, we get
\begin{equation*}
B,_a=0,\quad Y(M)=\frac{\chi_4}{\chi_3}.
\end{equation*}
Now, we insert symmetry generators, quadratic form of
$f(\mathcal{G})$ model in Eq.(\ref{37}) with $B=\frac{\chi_1
r}{\chi_3}+\chi_8$, $M=r^2$, $e^b=\left(1-h(r)/r\right)^{-1}$,
$a(r)=-k/r$ and obtain a non-linear equation given by
\begin{eqnarray}\nonumber&&
2 h(r)^4 \chi_7 w(\{k (r (4 r+k (-1+4 r^2))+h(r)(k-5 r-8 k r^2+12
r^3+4 r h(r)\\\nonumber&&\times (k-3 r))+r^2(1-12 r^2+12 r h(r))
h'(r))\}\{r^3 (r-h(r))^2\}^{-1})^2 -8 r h(r)^3
\\\nonumber&&\times\chi_7 w(\{k (r (4 r+k (-1+4
r^2))+h(r)(k-5 r-8 k r^2+12 r^3+4 (k-3 r)\\\nonumber&&\times r
h(r))+r^2(1-12 r^2+12 r h(r))h'(r))\}\{r^3 (r-h(r))^2\}^{-1})^2 +12
r^2 h(r)^2\\\nonumber&&\times\chi_7 w(\{k(r(4 r+k(-1+4 r^2))+h(r)
(k-5 r-8 k r^2+12 r^3+4 (k-3 r)\\\nonumber&&\times r h(r))+r^2 (1-12
r^2+12 r h(r)) h'(r))\}\{r^3 (r-h(r))^2\}^{-1})^2-8 r^3
h(r)\\\nonumber&&\times\chi_7 w(\{k(r(4 r+k(-1+4 r^2))+h(r) (k-5 r-8
k r^2+12 r^3+4 (k-3 r)\\\nonumber&&\times rh(r))+r^2(1-12 r^2+12 r
h(r)) h'(r))\}\{r^3 (r-h(r))^2\}^{-1})^2+2 r^4\chi_7
w\\\nonumber&&(\{k(r(4 r+k(-1+4 r^2))+h(r)(k-5 r-8 k r^2+12 r^3+4
(k-3 r) r h(r))\\\nonumber&&+r^2(1-12 r^2+12 r h(r))h'(r))\}\{r^3
(r-h(r))^2\}^{-1})^2+\{-(k^2+3 k r\\\nonumber&&+24 r^2)
h(r)+r(-4+k^2+4 k r+32 r^2-r (k+8 r) h'(r))\}\{2
r^5\}^{-1}\\\label{40}&&+\rho_0 e^{k(1+w)/2 r w} r^{12}=0.
\end{eqnarray}

The numerical solution of this equation leads to analyze the
behavior of viability of quadratic $f(\mathcal{G})$ model,
geometrical properties of shape function, presence/absence of
ordinary and exotic matter graphically. In figure \textbf{5}, we
explore the consistency of quadratic model with standard models of
modified GB gravity. In both plots, the positively decreasing (left)
and increasing (right) curves preserve the viability constraints as
$f_{\mathcal{G}}>0$ and $f_{\mathcal{G}\mathcal{G}}>0$. In figure
\textbf{6}, we study the geometry of WH constructed by quadratic
$f(\mathcal{G})$ model and corresponding shape function. The left
plot demonstrates the asymptotically flat shape of WH as
$h/r\rightarrow0$ when $r\rightarrow\infty$. In the right plot, the
trajectory of $h(r)-r$ locates WH throat at $r_0=7$ and at this
point, the derivative of the shape function is found to be positive
but greater than 1. This analysis defines a horizon-free
asymptotically flat WH whose throat is located at $r_0=7$ and
$h(r_0)=r_0$ in the background of phantom energy ($w=-1$).

\begin{figure}\center
\epsfig{file=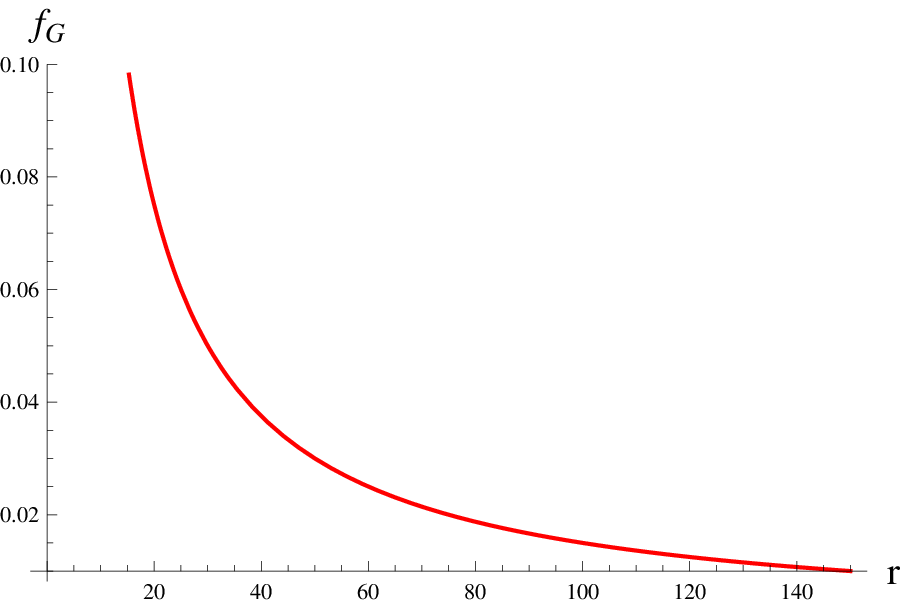,width=0.45\linewidth}
\epsfig{file=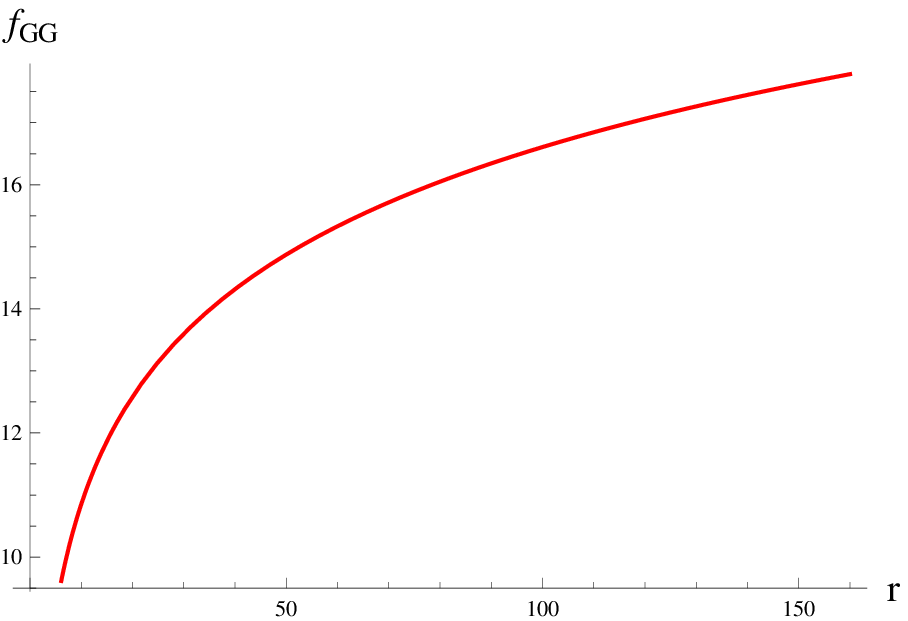,width=0.45\linewidth} \caption{Evolution of
quadratic $f(\mathcal{G})$ model versus $r$ for $\chi_7=-0.1$,
$\rho_0=-1.5$, $k=0.05$ and $w=-1$.}
\end{figure}
\begin{figure}\center
\epsfig{file=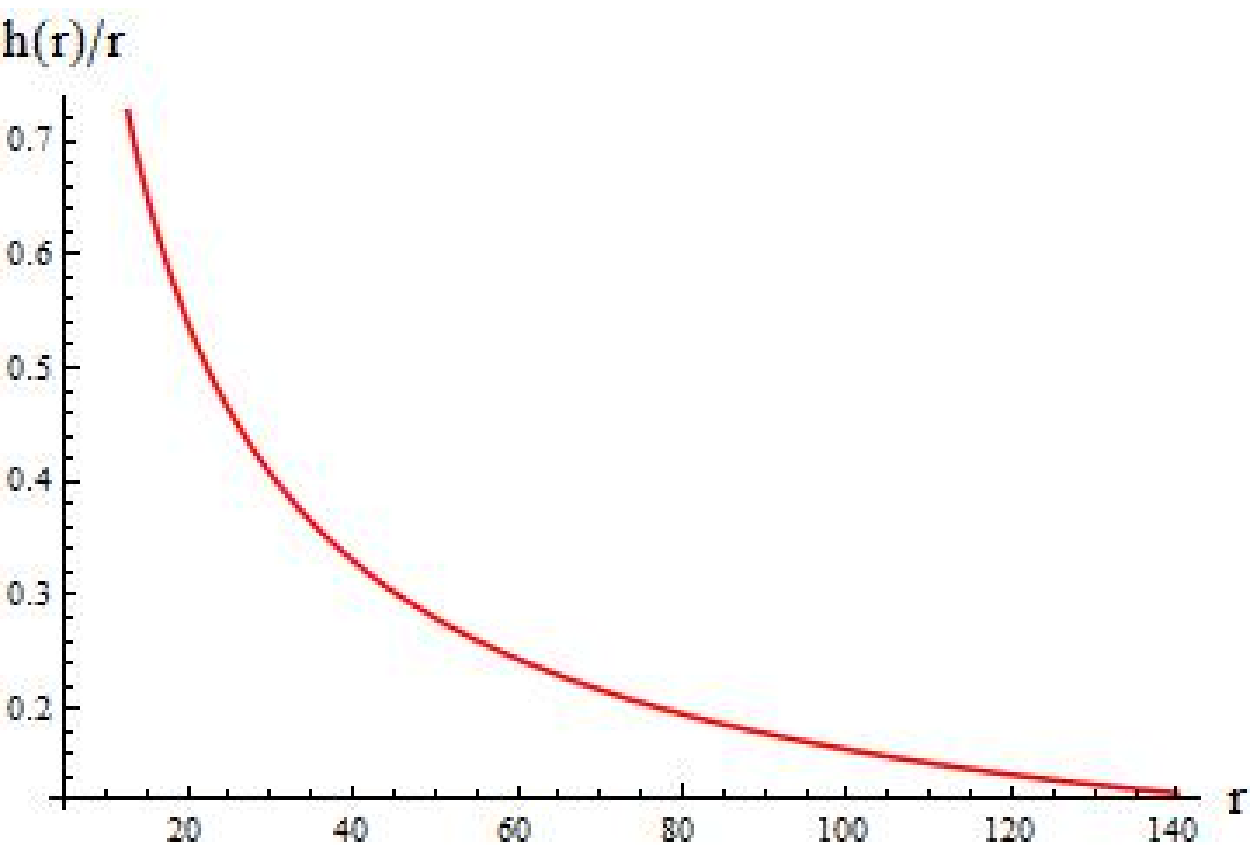,width=0.45\linewidth}
\epsfig{file=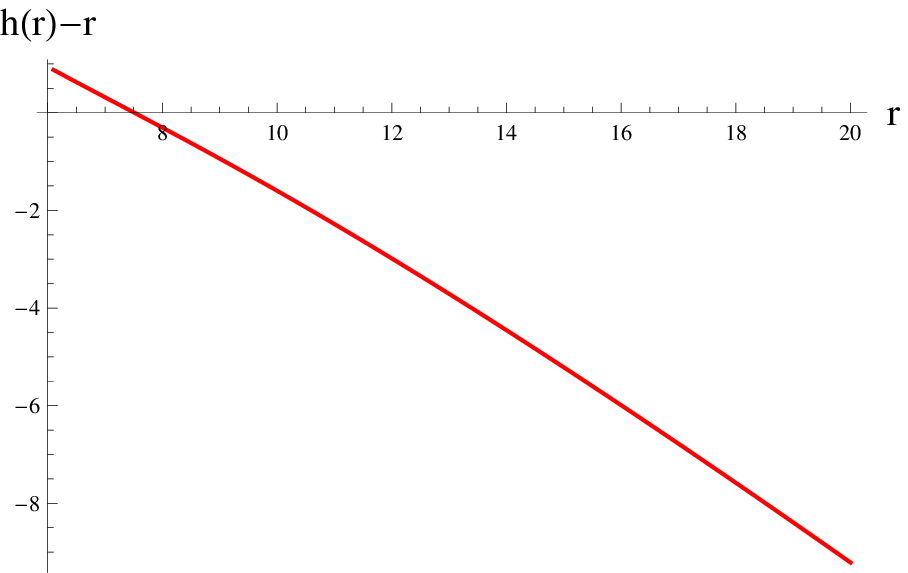,width=0.45\linewidth}\caption{Variation of the
shape function versus $r$.}
\end{figure}
\begin{figure}\center
\epsfig{file=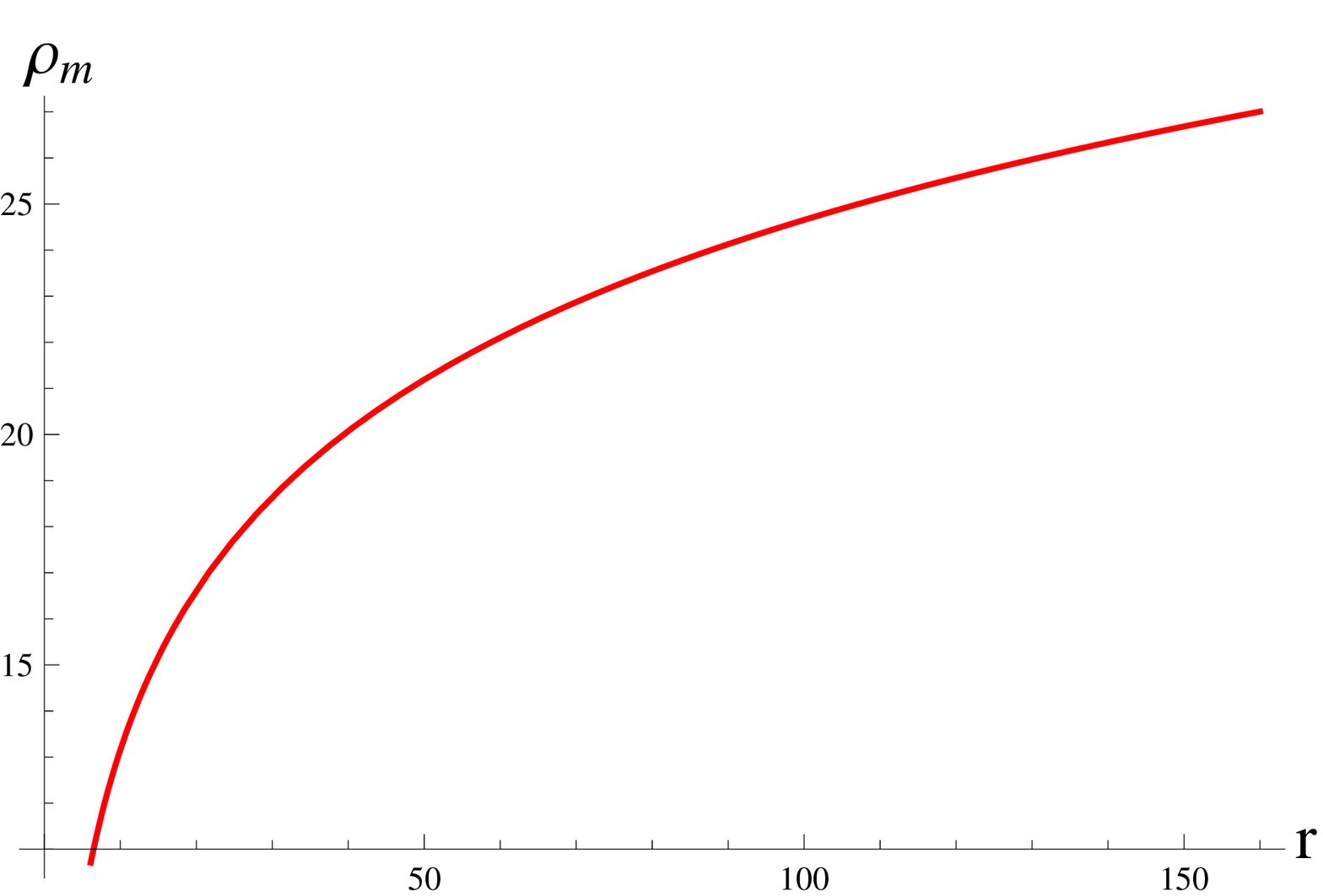,width=0.45\linewidth}\epsfig{file=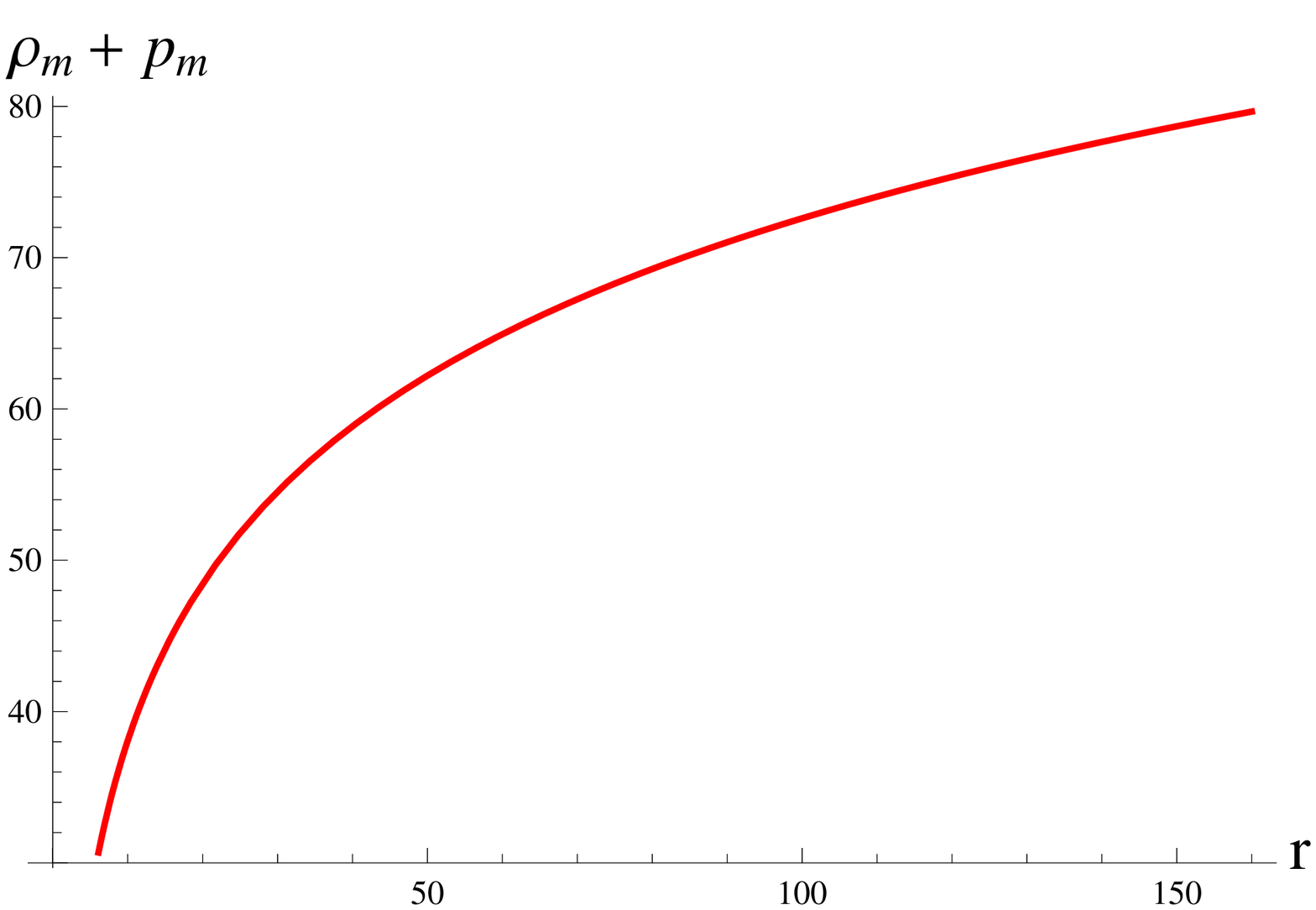,width=0.45\linewidth}
\caption{Evolution of energy bounds versus $r$ for $w=-1$.}
\end{figure}
\begin{figure}\center
\epsfig{file=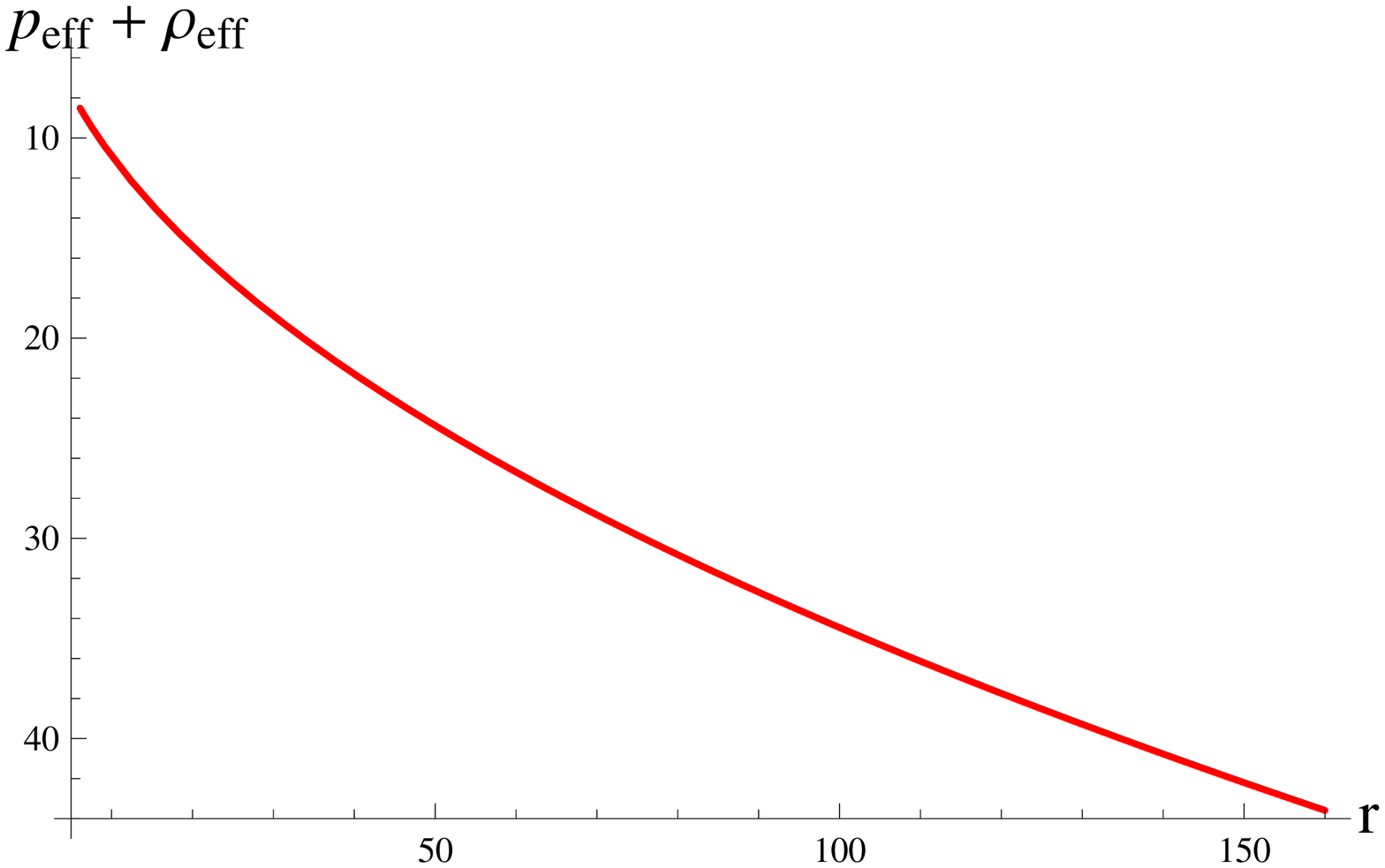,width=0.45\linewidth}\caption{Evolution of
effective NEC versus $r$ for $w=-1$.}
\end{figure}
Now, we examine physical and traversable behavior of WH via energy
conditions for ordinary matter and effective NEC, respectively.
Figure \textbf{7} explores the nature of matter variables. In both
plots, the matter variables are found to be increasing positively
ensuring that the presence of ordinary matter is confirmed as
$\rho_m\geq0$ and $\rho_m+p_m\geq0$. The behavior of effective NEC
versus $r$ is shown in figure \textbf{8}. The trajectory of
effective matter variables is found to be negative increasing as $r$
increases. This behavior indicates that at the throat, the effective
NEC is violated ensuring the presence of exotic matter leading to
traversable WH solution. In this regard, the realistic horizon-free
asymptotically flat WH solution admits traversable behavior for
quadratic $f(\mathcal{G})$ model.

\subsection*{Case III: $\beta=\delta\neq0$}

In this case, we solve the system of over determined equations
(\ref{30})-(\ref{36}) and  get
\begin{eqnarray*}\nonumber
\alpha&=&\phi_1+\phi_2e^{-a},\quad\gamma=M(\phi_3-\phi_4),
\quad\tau=\phi_5r+\phi_6,
\\\label{38b}f(\mathcal{G})&=&\phi_7e^{\phi_7\mathcal{G}+\phi_8}
+\phi_9,\quad\delta=\phi_3e^bT_1(M)T_2(\mathcal{G}),
\quad\beta=\phi_3e^{b/\phi_7}T_1(M)T_2(\mathcal{G}),
\end{eqnarray*}
where $\phi_{i's}$ are arbitrary constants and the explicit form of
$f$ defines exponential model of modifies GB gravity. Using above
solutions in Eq.(\ref{29}), we have
\begin{equation*}
B,_a=0,\quad T_1(M)=\phi_7e^{\phi_9M},\quad
T_2(\mathcal{G})=\phi_7e^{-\phi_{10}\mathcal{G}}.
\end{equation*}
Now, we insert symmetry generators, exponential $f(\mathcal{G})$
model, $B=\frac{\phi_1 r}+\frac{\phi_8}{\phi_3}$, $M=r^2$,
$e^b=\left(1-h(r)/r\right)^{-1}$ and $a(r)=-k/r$ in Eq.(\ref{37})
which leads to the following non-linear equation
\begin{eqnarray}\nonumber&&
e^{-\frac{k}{2 r}}+\sqrt{r/(r-h(r))} (e^{\frac{k (1+w)}{2 r
w}}\rho_0 (-\frac{\phi_3e^{k/r}}{2 w}+\frac{1}{2} \phi_7 \phi_9
\phi_{10} \exp\{r^2-((r-h(r))^2\\\nonumber&&\times2[(k(r(4 r+k (-1+4
r^2))+h(r)(k-5 r-8 k r^2+12 r^3+4 (k-3 r) r
h(r))\\\nonumber&&+r^2(1-12 r^2+12 r h(r))h'(r)))\{r^3
(r-h(r))^2\}^{-1}])r^{-6}\}(r-h(r)/r))\\\nonumber&& +\{2r^{-5}
\phi_7 \phi_9 \phi_{10}e^{\phi_1+r^2}(-r+h(r))[\{k(r(4 r+k (-1+4
r^2))+h(r)(k-5 r\\\nonumber&&-8 k r^2+12 r^3+4 (k-3 r) r
h(r))+r^2(1-12 r^2+12 r h(r))
h'(r))\}\{r^3\\\nonumber&&\times(r-h(r))^2\}^{-1}]\}
+(\phi_3e^{k/r}/2-\{\phi_7 \phi_9 \phi_{10}\exp\{r^2-(2
(r-h(r))^2[\{k (r(4 r\\\nonumber&&+k(-1+4 r^2))+h(r)(k-5 r-8 k
r^2+12 r^3+4 (k-3 r) r h(r))+r^2(12 r h\\\nonumber&&-12
r^2-1)h'(r))\}\{r^3 (r-h(r))^2\}^{-1}])r^{-6}\} r\}\{2
(-r+h(r))\}^{-1})(\exp\{\phi_1\\\nonumber&&+(2 (r-h(r))^2[\{k(r(4
r+k(-1+4 r^2))+h(r) (k-5 r-8 k r^2+12 r^3+4r\\\nonumber&&\times(k-3
r)h(r))+r^2(1-12 r^2+12 r h(r)) h'(r))\}\{r^3
(r-h(r))^2\}^{-1}])r^{-6}\}\\\nonumber&&-\{2 \exp\{\phi_1+(2
(r-h(r))^2[\{k(r(4 r+k(-1+4 r^2))+h(r) (k-5 r-8 k
r^2\\\nonumber&&+12 r^3+4 (k-3 r) r h(r))+r^2(1-12 r^2+12 r h(r))
h'(r))\}(r^3 (r-h(r))^2)^{-1}])\\\nonumber&&\times
r^{-6}\}(-r+h(r))^2[\{k(r(4 r+k(-1+4 r^2))+h(r) (k-5 r-8 k r^2+12
r^3\\\nonumber&&+4 (k-3 r) r h(r))+r^2(1-12 r^2+12 r h(r))
h'(r))\}(r^3 (r-h(r))^2)^{-1}]\}r^{-6}\\\nonumber&&+((k^2+3 k r+24
r^2)h(r)+r(4-k^2-4 k r-32 r^2+r (k+8 r)
h'(r)))\\\label{41}&&\times\{2 r^5\}^{-1}))r^2=0.
\end{eqnarray}

The numerical solution of this equation leads to study viability of
exponential $f(\mathcal{G})$ model and geometry of WH configuration.
We also establish graphical analysis to explore the exotic/ordinary
nature of matter that defines physically acceptable and traversable
WH configuration. In figure \textbf{9}, we discuss the viable
behavior of exponential model of modified GB gravity. In both plots,
the positively decreasing (left) and increasing (right) curves show
that $f_{\mathcal{G}}>0$ and $f_{\mathcal{G}\mathcal{G}}>0$ implying
consistency with viable GB models. Figure \textbf{10} elaborates the
geometry of numerically constructed WH. In the left plot, positively
decreasing curve follows asymptotically flat shape as
$r\rightarrow\infty$. The right plot locates WH throat at $r_0=12$
and at this point, the derivative of the shape function remains
greater than 1. This analysis defines horizon-free asymptotically
flat WH that possesses a throat at $r_0=12$ such that $h(r_0)=r_0$.
\begin{figure}\center
\epsfig{file=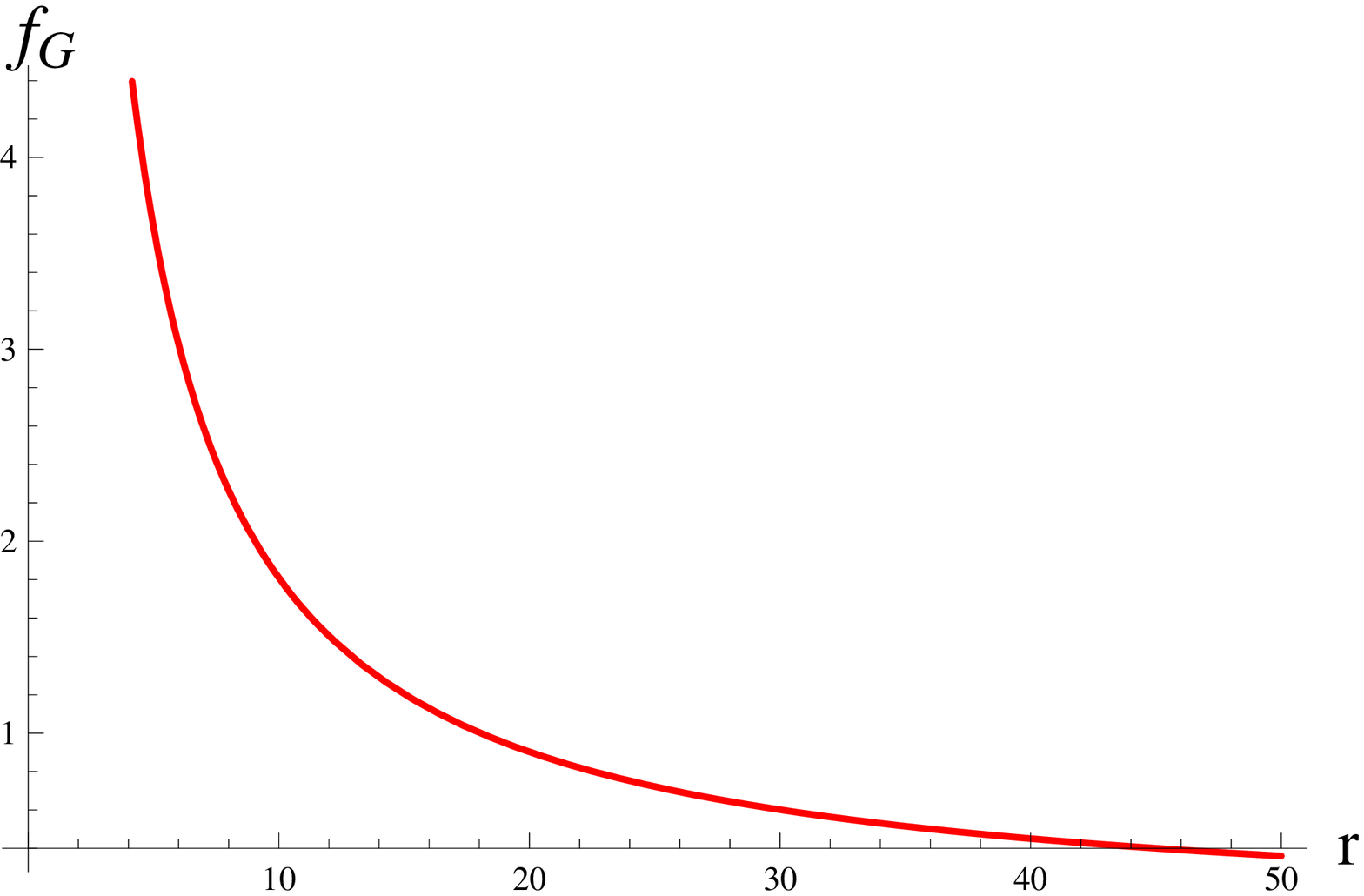,width=0.45\linewidth}
\epsfig{file=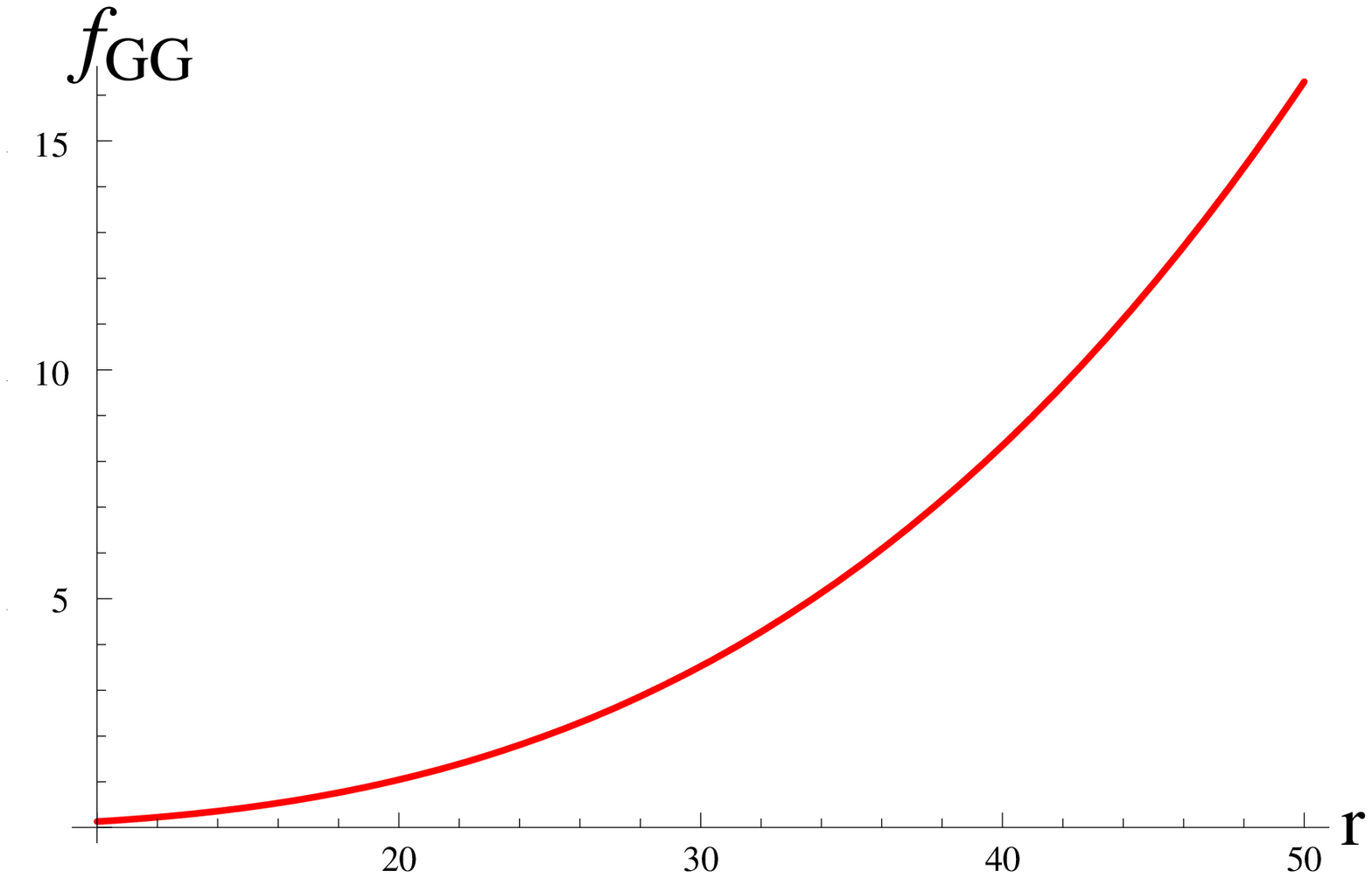,width=0.45\linewidth} \caption{Evolution of
$f(\mathcal{G})$ model versus $r$ for $\phi_1=0.01$, $\phi_7=-1.25$,
$\phi_9=1$, $\phi_{10}=1.5$, $k=0.5$, $w=-1$, and $\rho_0=-1.5$.}
\end{figure}
\begin{figure}\center
\epsfig{file=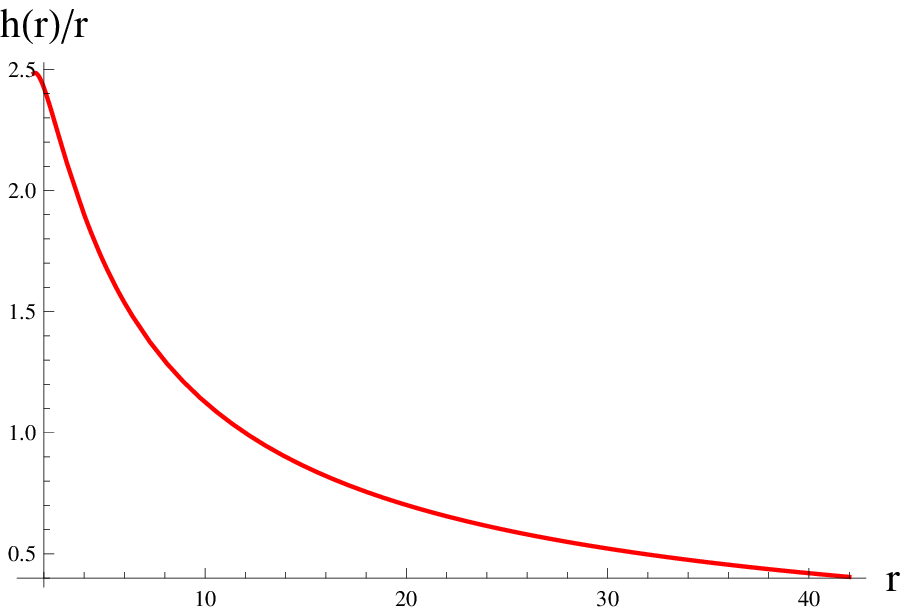,width=0.45\linewidth}
\epsfig{file=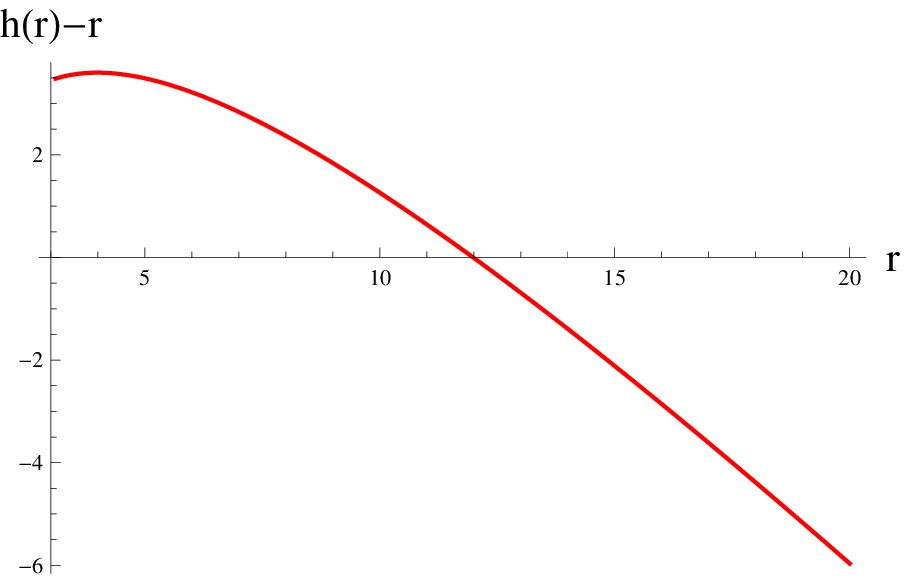,width=0.45\linewidth}\caption{Variation of the
shape function versus $r$.}
\end{figure}

To explore the existence of physically viable and traversable WH, we
establish graphical analysis of ordinary as well as exotic matter
variables in figures \textbf{11} and \textbf{12}. In plots of
\textbf{11}, the trajectories are found to be positively increasing
justifying the existence of realistic WH supported by ordinary
matter inside the throat. Figure \textbf{12} shows the traversable
behavior of WH due to violation of effective NEC that introduces
repulsive effects into the WH throat. In case of exponential
$f(\mathcal{G})$ model, the realistic as well as traversable WH
exists in the background of accelerating cosmos.
\begin{figure}\center
\epsfig{file=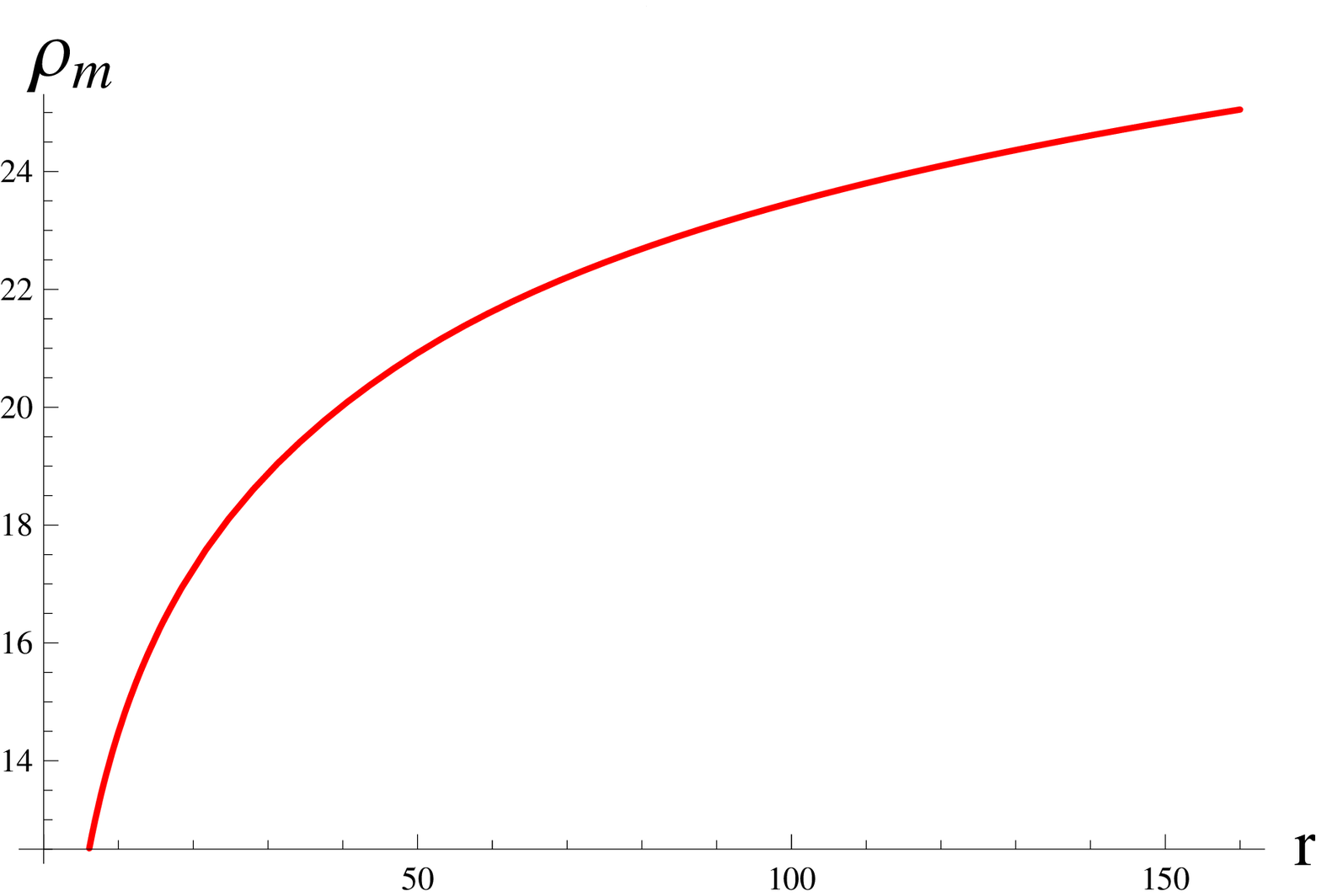,width=0.45\linewidth}
\epsfig{file=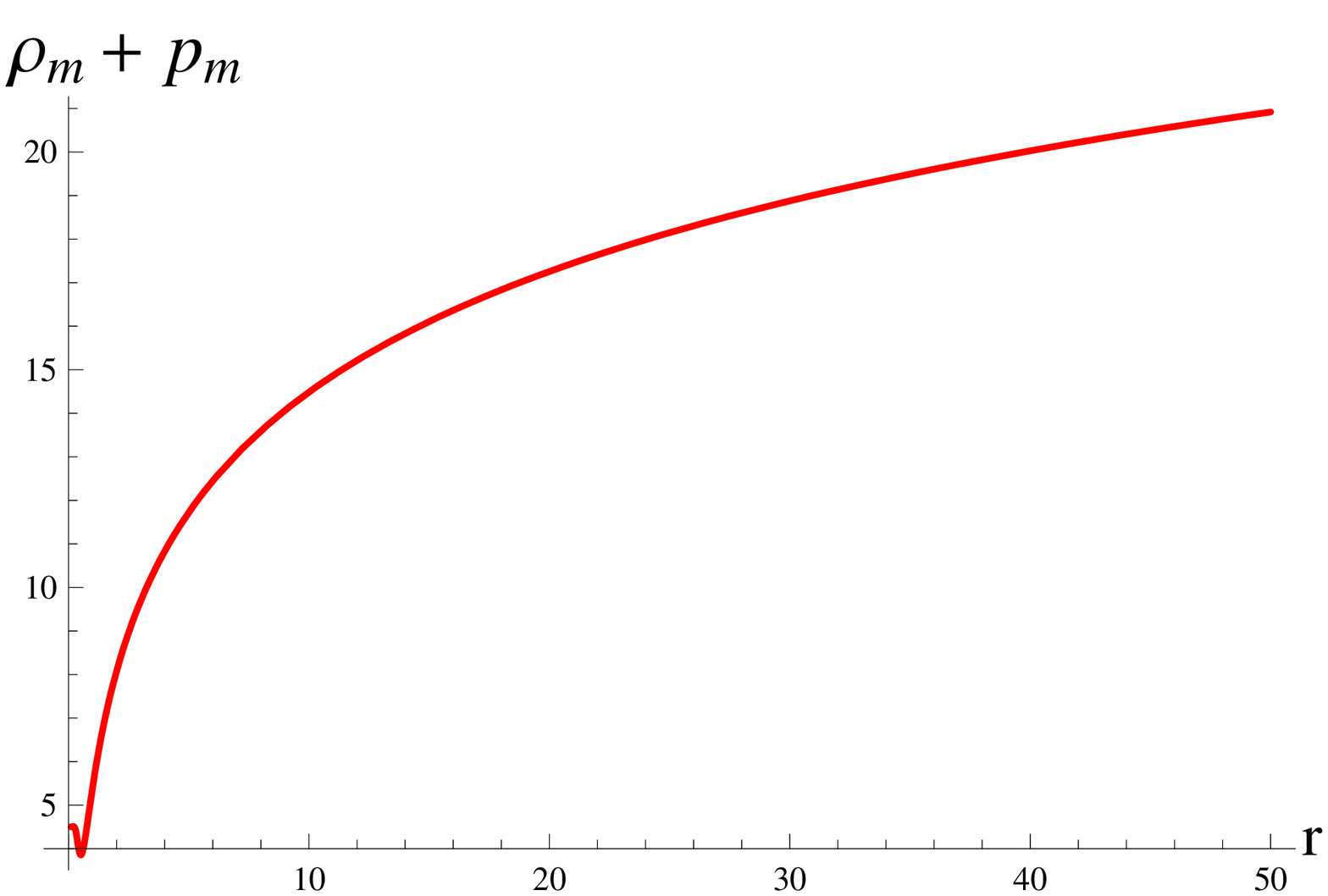,width=0.45\linewidth}\caption{Evolution of
energy bounds versus $r$ for $w=-1$.}
\end{figure}
\begin{figure}\center
\epsfig{file=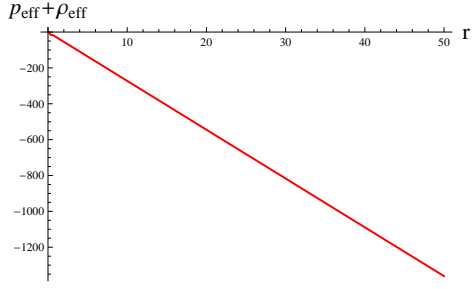,width=0.45\linewidth} \caption{Evolution of
effective NEC versus $r$ for $w=-1$.}
\end{figure}

\section{Stability Analysis}

Here, we examine the stability of WH solutions through
Tolman-Oppenheimer-Volkoff (TOV) equation for linear, quadratic and
exponential $f(\mathcal{G})$ models in the context of accelerated
($w=-1$) as well as decelerated ($w=0.3$) expanding cosmos. For
perfect fluid configuration, the radial function of Bianchi identity
$(\nabla_\alpha T^{\alpha\beta}=0)$ characterizes TOV equation as
\begin{equation}\label{56}
\frac{a'}{2}\left(p_m+\rho_m\right)+\frac{dp_m}{dr}=0.
\end{equation}
The divergence of stress-energy tensor with respect to modified
terms and Eq.(\ref{56}) leads to define modified TOV equation given
by
\begin{equation}\label{57}
p'_{eff}+\mathcal{M}_{eff}(p_{eff}+\rho_{eff})+\frac{M'}{M}\left(T_{11}^c
-\frac{T_{22}^ce^b}{M}\right)=0,
\end{equation}
where $p_{eff}=T_{11}^{(c)}+p_m$, $\rho_{eff}=T_{00}^{(c)}+\rho_m$
and $\mathcal{M}_{eff}=\frac{a'e^{b-a}}{2}$ defines effective
gravitational mass. The expressions for gravitational
$\mathcal{F}_g$ and hydrostatic $\mathcal{F}_h$ forces can be
written as
\begin{eqnarray}\nonumber
\mathcal{F}_h&=&\frac{d}{dr}(T_{11}^{(c)}+p_m),\\\nonumber
\mathcal{F}_g&=&\mathcal{M}_{eff}\left(p_{eff}+\rho_{eff}\right)
+\frac{M'}{M}\left(T_{11}^c-\frac{T_{22}^ce^b}{M}\right).
\end{eqnarray}
\begin{figure}\center
\epsfig{file=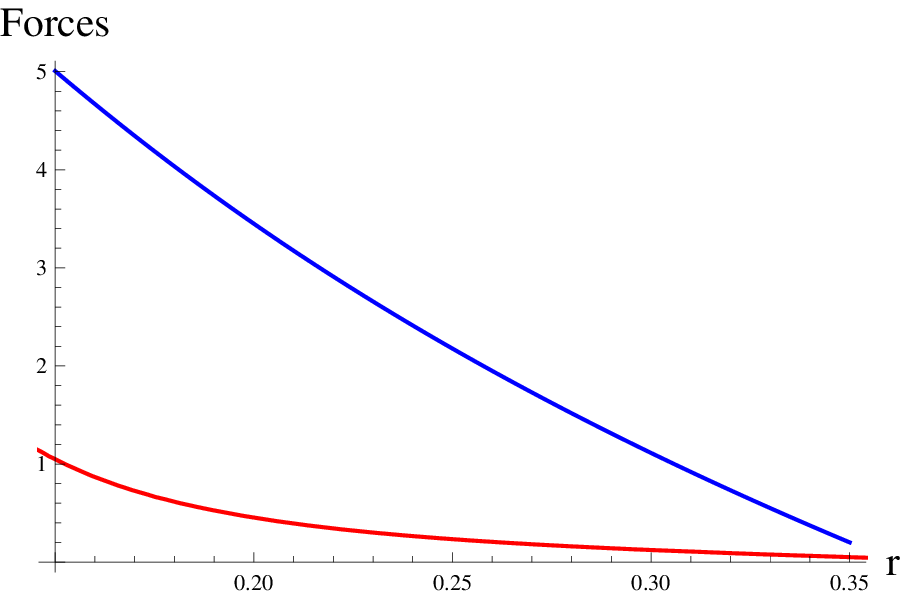,width=0.4\linewidth}
\epsfig{file=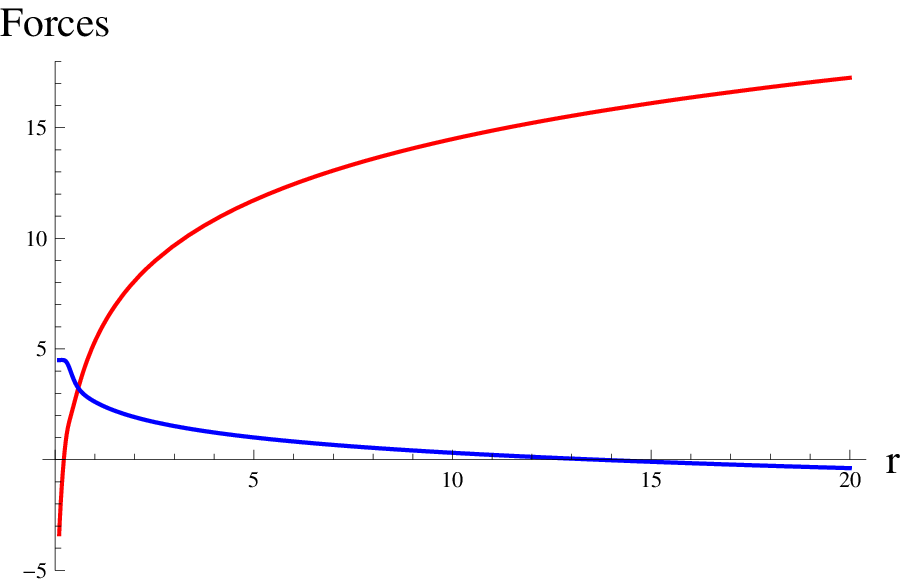,width=0.4\linewidth}\caption{Plots of
$\mathcal{F}_h$ (blue) and $\mathcal{F}_g$ (red) versus $r$ for
$f(\mathcal{G})=\xi_1\mathcal{G}+\xi_2$ (left) and
$f(\mathcal{G})=\phi_7e^{\phi_7\mathcal{G}+\phi_8}+\phi_9$ (right),
$w=-1$.}
\end{figure}
These dynamical forces significantly explore the stable/unstable
state of static configuration. Here, we discuss the
stability/instability of static traversable and physically viable WH
solutions corresponding to linear, quadratic and exponential
$f(\mathcal{G})$ models. The stable WH may exists if these dynamical
forces counterbalance each other effect, i.e.,
$\mathcal{F}_h+\mathcal{F}_g=0$ or $\mathcal{F}_g$=-$\mathcal{F}_h$.
\begin{figure}\center
\epsfig{file=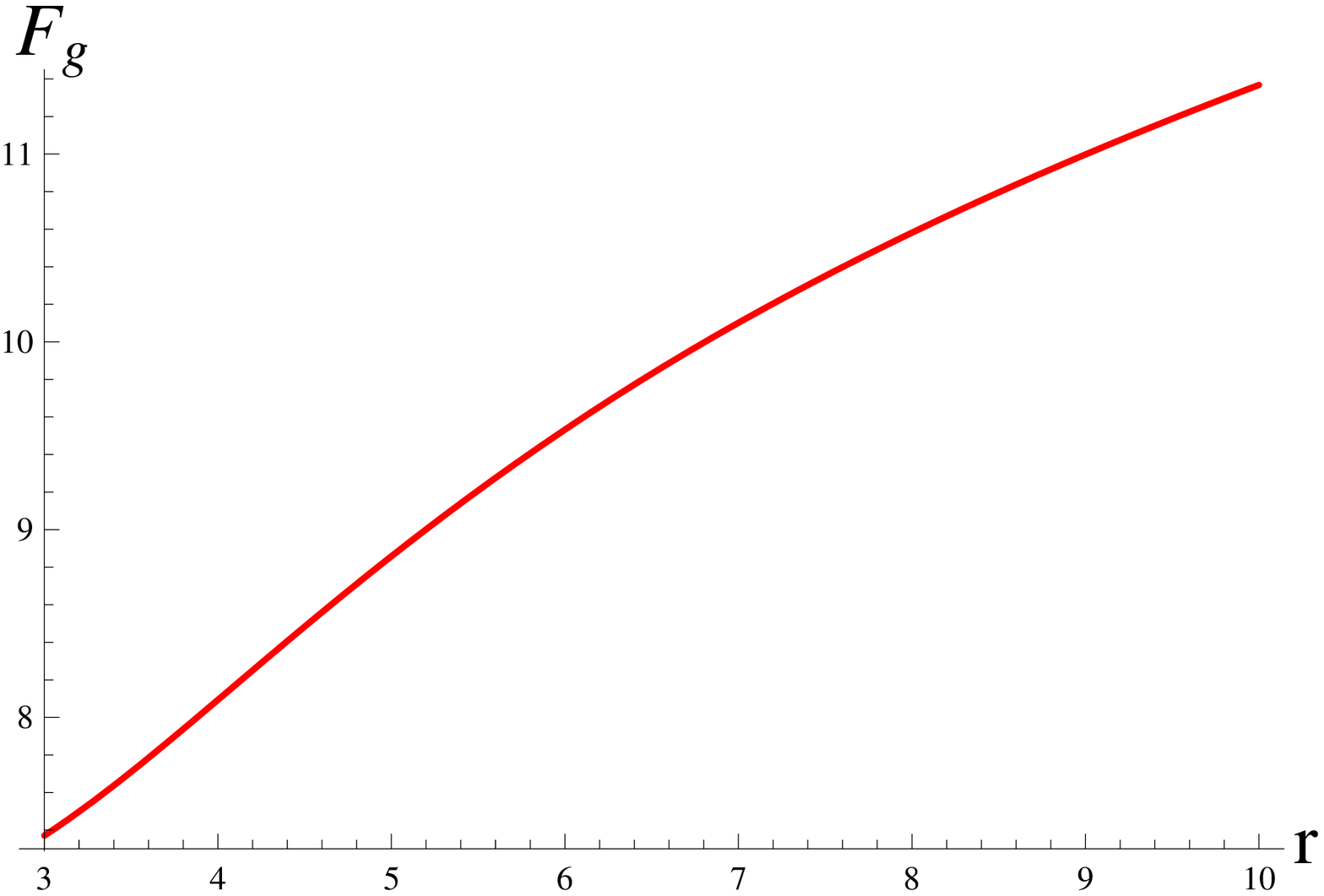,width=0.4\linewidth}
\epsfig{file=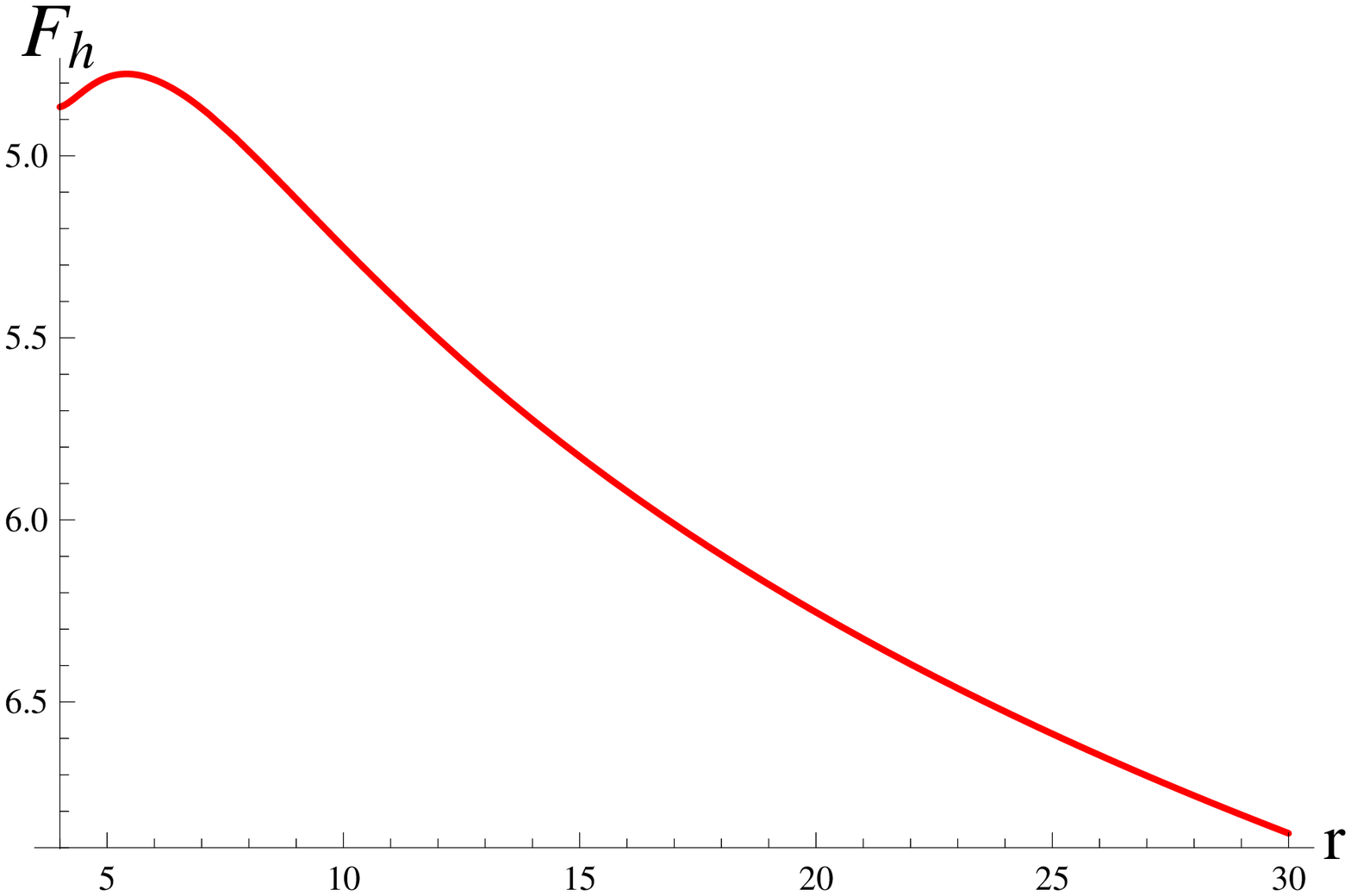,width=0.4\linewidth}\caption{Plots of
$\mathcal{F}_g$ and $\mathcal{F}_h$ versus $r$ for
$f(\mathcal{G})=\chi_7\mathcal{G}^2+\chi_8\mathcal{G}+\chi_1$ with
$w=-1$.}
\end{figure}

Figure \textbf{13} shows the behavior of gravitational and
hydrostatic forces for both linear (left plot) as well as
exponential (right plot) models in the context of accelerated cosmos
($w=-1$). In the left plot, the trajectories corresponding to
hydrostatic and gravitational forces are found to be positively
decreasing leading to stable state of WH solution due to null effect
of these forces. The analysis of right plot indicates that the
stable state of WH solution can be achieved as gravitational and
hydrostatic forces are evolving positively but in opposite direction
and consequently, canceling the effects of each other. In figure
\textbf{14}, we study the stable state of WHs for quadratic GB model
when universe experiences accelerated phase of expansion ($w=-1$).
This analysis indicates that the horizon-free asymptotically flat
traversable and physically viable WHs are stable against accelerated
expanding cosmos for both quadratic as well as exponential models of
$f(\mathcal{G})$ gravity.

\section{Final Remarks}

In Einstein's gravity, the violation of NEC is the basic requirement
for the existence of traversable WH. The violation of NEC defines
exotic nature of matter that should be minimized for a physically
viable WH. For modified theories, the stress-energy tensor relative
to ordinary matter fulfills energy bounds ensuring the presence of a
viable WH while the existence of exotic matter is confirmed by the
effective matter variables which do not obey energy bounds like
effective NEC. In this paper, we have used Noether symmetry
technique to evaluate some exact solutions that helps to construct
static WHs in $f(\mathcal{G})$ theory. We have discussed the
presence of exotic and normal matter in WHs through effective and
ordinary energy bounds. We have also examined stable/unstable state
of constructed WHs via modified TOV equation.

We have used the invariance condition to solve over determined
system of equations and evaluated symmetry generator, related
conserved quantities and three different $f(\mathcal{G})$ models
such as linear, quadratic and exponential models. In the context of
these models, we have formulated WH solutions in the background of
accelerated expanding cosmos ($w=-1$) and analyzed the WH geometry
for variable red-shift function $a(r)=-k/r$. For linear
$f(\mathcal{G})$ model, we have found horizon-free WH which is found
to be asymptotically flat in a very short interval of $r$. The
throat of this WH is located at $r_0=0.34$ with $h'(r_0)<1$ implying
flaring-out condition is preserved. For both quadratic and
exponential models, the WH geometry is compatible with
Morris-Thorne's suggested geometry, i.e., the finite red-shift
function introduces horizon-free $(h(r)<r)$, asymptotically flat WH
as $r\rightarrow\infty$ while flaring-out condition violates
($h(r_0)=r_0$ but $h'(r_0)>1$) for both models. Using numerical
solution of shape function, the viability of new $f(\mathcal{G})$
models is examined graphically. The graphical interpretation
indicates that the derivative of $f(\mathcal{G})$ models are
positive ensuring viable state of these models.

A WH is traversable if there exists strong repulsive effects or
exotic matter near WH throat while physically viable WH is defined
by ordinary matter. The violation of effective NEC
($p_{eff}+\rho_{eff}\leq0$) confirms the presence of repulsive force
inside the throat. The positivity of matter variables like
$\rho_m\geq0$, $\rho_m+p_m\geq0$, $\rho_m-p_m\geq0$ and
$\rho_m+3p_m\geq0$ preserve consistency with energy conditions,
i.e., NEC, WEC, DEC and SEC relative to ordinary matter and
consequently, supports physically viable WH. For all formulated
$f(\mathcal{G})$ models, the violation of effective NEC inside WH
throat confirms the presence of traversable WH while fulfillment of
ordinary bounds leads to physically viable WHs in the background of
accelerated expansion.

The stability/instability of these traversable and physically viable
WHs is examined via modified TOV equation. For linear
$f(\mathcal{G})$ model, the WH configuration surrounded by
accelerated expanding cosmos is found to be unstable due to
unbalanced state of hydrostatic and gravitational forces. In case of
quadratic and exponential models with $w=-1$, the WH solutions
preserves equilibrium state as the dynamical forces counterbalance
each other effect. Sharif and Nawazish \cite{20} have constructed
traversable and realistic WH solution in $f(R)$ gravity for constant
as well as variable forms of red-shift function. They have
formulated exponential form of $f(R)$ and also considered a standard
power-law $f(R)$ models. The stability analysis of both models
indicates that WH solutions are stable when universe experiences
decelerated rate of expansion while in the presence of accelerated
expansion, these configurations become unstable. In the present
work, we have evaluated three viable $f(\mathcal{G})$ models, i.e.,
linear, quadratic and exponential models that yield traversable and
physically viable WHs. For quadratic and exponential models, these
configurations are stable whereas in case of linear model, this
stability is disturbed in the presence of phantom energy.

\end{document}